\documentclass[fleqn,usenatbib]{mnras}

\usepackage{newtxtext,newtxmath}

\usepackage[T1]{fontenc}
\usepackage{ae,aecompl}

\usepackage{graphicx}	
\usepackage{amsmath}	
\usepackage{amssymb}	
\usepackage{float}
\usepackage{caption}
\usepackage{mathtools}
\usepackage{breqn}
\usepackage[english]{babel}
\usepackage{blindtext}

\title[The Impact of Fiducial Cosmology on BAO Inference]{The Impact of the Fiducial Cosmology Assumption on BAO Cosmological Parameter Inference}

\author[P. Carter et al.]{Paul Carter$^{1}$\thanks{E-mail: paul.carter1@port.ac.uk},
Florian Beutler$^{1, 2}$,
Will J. Percival$^{3, 4}$,
Joseph DeRose$^{5, 6, 7}$,
\newauthor Risa H. Wechsler$^{5, 6, 7}$,
Cheng Zhao$^{8}$
\\
$^{1}$Institute of Cosmology \& Gravitation, University of Portsmouth, Dennis Sciama Building, Portsmouth, PO1 3FX, UK\\
$^{2}$Lawrence Berkeley National Lab, 1 Cyclotron Rd, Berkeley CA 94720, USA\\
$^{3}$Waterloo Centre for Astrophysics, Department of Physics and Astronomy, University of Waterloo, Waterloo, ON N2L 3G1, Canada\\
$^{4}$Perimeter Institute for Theoretical Physics, Waterloo, ON N2L 2Y5, Canada\\
$^{5}$Department of Physics, Stanford University, 382 Via Pueblo Mall, Stanford, CA 94305, USA\\
$^{6}$Kavli Institute for Particle Astrophysics \& Cosmology, P. O. Box 2450, Stanford University, Stanford, CA 94305, USA\\
$^{7}$SLAC National Accelerator Laboratory, Menlo Park, CA 94025, USA\\
$^{8}$Laboratory of Astrophysics, Ecole Polytechnique F\'ed\'erale de Lausanne (EPFL), Observatoire de Sauverny, 1290 Versoix, Switzerland\\
}

\date{Accepted XXX. Received YYY; in original form ZZZ}

\pubyear{2019}

\begin{document}
\label{firstpage}
\pagerange{\pageref{firstpage}--\pageref{lastpage}}
\maketitle

\begin{abstract}
Standard analysis pipelines for measurements of Baryon Acoustic Oscillations (BAO) in galaxy surveys make use of a fiducial cosmological model to guide the data compression required to transform from observed redshifts and angles to the measured angular and radial BAO peak positions. In order to remove any dependence on the fiducial cosmology from the results, all models compared to the data should mimic the compression and its dependence on the fiducial model. In practice, approximations are made when testing models: (1) There is assumed to be no residual dependence on the fiducial cosmology after reconstruction, (2) differences in the distance--redshift relationship are assumed to match a linear scaling, and (3) differences in clustering between true and fiducial models are assumed to be removed by the free parameters used to null the non-BAO signal. We test these approximations using the current standard measurement procedure with a set of halo catalogs from the {\sc Aemulus} suite of $N$-body simulations, which span a range of $w\mathrm{CDM}$ cosmological models. We focus on reconstruction of the primordial BAO and locating the BAO. For the range of $w\mathrm{CDM}$ cosmologies covered by the {\sc Aemulus} suite, we find no evidence for systematic errors in the measured BAO shift parameters $\alpha_{\parallel}$ and $\alpha_{\bot}$ to $< 0.1\%$. However, the measured errors $\sigma_{\alpha_{\parallel}}$ and $\sigma_{\alpha_{\bot}}$ show a notable absolute increase by up to $+0.001$ and $+0.002$ respectively in the case that the fiducial cosmology does not match the truth. These effects on the inferred BAO scale will be important given the precision of measurements expected from future surveys including DESI, Euclid, and WFIRST.
\end{abstract}




\section{Introduction}
\label{sec:intro}

The measurement of the projected Baryon Acoustic Oscillation (BAO) signal in galaxy surveys has become an essential probe of cosmology \citep{2017MNRAS.470.2617A}. Prior to recombination, the temperature of the Universe is higher than the ionisation energy of electrons. Baryonic matter and radiation are coupled in a plasma state. This photon--baryon fluid acts under gravitational forces around density perturbations and also under radiation pressure. The interplay between these opposing forces generates acoustic oscillations in the fluid until photons decouple fully at $z \sim 1020$. The imprint of the BAO is left in overdensity peaks at $r_{d} \sim 150\mathrm{Mpc}$ in the two-point statistics of the matter field. Galaxies eventually form in regions of higher overdensity and hence act as biased tracers of the matter field on large scales, and therefore also reveal the BAO. In turn, the BAO signal can be used as a standard ruler to constrain the distance--redshift relation.

The BAO peak has been observed in many galaxy samples, with the first observations in the Sloan Digital Sky Survey (SDSS; \citealt{2000AJ....120.1579Y, 2005ApJ...633..560E}) and the 2-degree Field Galaxy Redshift Survey (2dFGRS; \citealt{2001MNRAS.328.1039C, 2001MNRAS.327.1297P, 2005MNRAS.362..505C}). Subsequently, the BAO peak has been detected in later SDSS data releases \citep{2010MNRAS.401.2148P, 2010ApJ...710.1444K}, the 6dFGS \citep{2011MNRAS.416.3017B, 2018MNRAS.481.2371C}, WiggleZ \citep{2011MNRAS.415.2892B}, BOSS \citep{2017MNRAS.464.3409B, 2017MNRAS.464.1168R, 2017MNRAS.470.2617A}, eBOSS Luminous Red Galaxies \citep{2017arXiv171208064B} and quasar samples \citep{2018MNRAS.473.4773A} and at higher redshift using the Ly-$\alpha$ forest in BOSS \& eBOSS \citep{2013JCAP...04..026S, 2014JCAP...05..027F, 2015A&A...574A..59D, 2019arXiv190403400D, 2019arXiv190403430B}.

The BAO has also been detected in the higher order statistics of the 3-point correlation function \citep{2017MNRAS.469.1738S} and bispectrum \citep{2018MNRAS.478.4500P} for the BOSS CMASS sample. The BAO feature has been measured using voids as the clustering tracer \citep{2016PhRvL.116q1301K, 2016MNRAS.459.4020L}, also in BOSS. These measurements have provided distance constraints that span from $z = 0$ out to $z \sim 0.8$ using conventional galaxy redshift surveys, and extend to $z \sim 1.5$ through eBOSS quasars and to $z \sim 2.3$ when including Ly-$\alpha$.
 
Recent measurements of the BAO position use density field reconstruction to sharpen the signal. \citet{2007ApJ...664..675E} proposed that, as the bulk flows that smear the acoustic peak are sourced from the density field potential itself, the galaxy map can itself be used to estimate the displacement field. Removal of these shifts can reduce the damping of the BAO and increase the $S/N$ of this feature. This increased $S/N$ results from information that was absorbed into higher order statistics being moved back into linear fluctuations \citep{2015PhRvD..92l3522S, 2017MNRAS.469.1738S}. Density field reconstruction has been applied in analyses of the 6dFGS \citep{2018MNRAS.481.2371C}, SDSS \citep{2012MNRAS.427.2132P,2015MNRAS.449..835R}, WiggleZ \citep{2014MNRAS.441.3524K} and throughout BOSS \citep{2017MNRAS.470.2617A}. These studies use either a  perturbation-theory-based approach that relies on the finite difference method \citep{2009PhRvD..79f3523P, 2009PhRvD..80l3501N}, or an alternative FFT-based iterative algorithm \citep{2014MNRAS.445.3152B, 2015MNRAS.453..456B}.
 
The level of statistical errors on the measurements listed above has, in general, been at the $\gtrsim 1\%$ level, with BOSS providing the currently best constraints with 1\% errors on the isotropic BAO scale \citep{2017MNRAS.470.2617A}. In the near future, multiple redshift surveys including DESI \citep{2016arXiv161100036D}, the European Space Agency Euclid mission \citep{2011arXiv1110.3193L}, and WFIRST \citep{2019arXiv190401174D} will provide sub-percent BAO-position errors in many redshift bins, providing unprecedented precision in the evolutionary history of the late-Universe and the cosmological model. The ability to access information about the BAO scale at this level of statistical errors means that a thorough understanding of systematic errors is required.

The standard BAO-measurement procedure adopted by recent galaxy survey analyses uses a fiducial cosmological model to guide the data compression required to transform from observed redshifts and angles to the measured angular and radial BAO peak positions. In order to remove any dependence on the fiducial cosmology from the final measurements, all models tested against the data need to also include the effects of data compression and its dependence on the fiducial model. This dependence is present in the reconstruction, power spectrum generation, and model fitting steps. In practice, a number of approximations are made:
\begin{enumerate}
    \item There is assumed to be no residual dependence on the fiducial cosmology after reconstruction.
    \item Differences in the distance--redshift relationship between true and fiducial cosmology are assumed to match a linear scaling.
    \item Differences in comoving clustering between true and fiducial cosmology are assumed to be removed by the same set of free parameters used to null the non-BAO signal in the correlation function or power spectrum.
\end{enumerate}
\noindent These dependencies have the potential to add systematic errors to BAO measurements.

These potential systematics were investigated in \cite{2018MNRAS.477.1153V} for the BOSS analysis. In the BOSS study, the effects of differences between fiducial and true cosmology were tested using the entire pipeline (reconstruction + power spectrum generation + BAO template fitting) for the simple case of deviations in $\Omega_{m}$ by 0.5\% within the $\Lambda \mathrm{CDM}$ model. \cite{2018MNRAS.479.1021D} provide further theoretical tests for future BAO measurements by using simulations designed to mitigate sample variance. Their focus was on model fitting but they covered other aspects, and they provide tests on systematics at the $\sim 0.01\%$ level. Theoretical work on extending the analytical framework for reconstruction in the case of the assumed cosmology being different from the true cosmology has been conducted in \cite{2019JCAP...02..027S}; here they find that under an assumption of linear theory there are negligible systematic errors $\Delta\alpha \sim 10^{-4}$ on the BAO position up to percent level changes in the full shape of the monopole and up to $5\%$ in the quadrupole.

In this paper, we present the results of analysing the {\sc Aemulus} suite \citep{2018arXiv180405865D} of simulations, which have cosmologies sampled from previous CMB likelihoods. We measure the BAO in halo catalogues drawn from each simulation, each analysed 40 times assuming a fiducial cosmology sampled from the same set of models. We investigate both the level of systematic errors that appear through the assumption of a fiducial cosmology for reconstruction and also when extracting the BAO scale through template fitting. Given that our simulations are timeslices, the linear scaling of the distance--redshift relationship between models will hold perfectly, so we cannot test evolution effects. This work expands the practical investigation of potential biases, extending the range of models tested and providing results that can be compared to theoretical work (e.g. \citealt{2019JCAP...02..027S}).

Our paper is organised as follows: Section~\ref{sec:sims} describes both the suite of $w\mathrm{CDM}$ {\sc Aemulus} simulations and also MD-PATCHY halo catalogues that have been used throughout. Section~\ref{sec:method} gives the methodology of the power spectrum, covariance matrix, and density field reconstruction techniques. The results are outlined, presenting the level of systematic errors measured due to the fiducial cosmology assumed during reconstruction only in Section~\ref{sec:resRec} and also including the model fitting in Section~\ref{sec:resFit}. We summarise our results in Section~\ref{sec:conclusion}.

\section{Simulations}  \label{sec:sims}

To test the dependence of measured BAO constraints on the fiducial cosmology assumed in the analysis and the size of the offset from the true cosmology, we measure BAO from a suite of 40 {\sc Aemulus} $w\mathrm{CDM}$ simulations \citep{2018arXiv180405865D}, each run assuming a different cosmological model. As this set is too small to be directly used to construct a covariance matrix, we use a set of 4096 halo catalogues drawn from MD-PATCHY simulations \citep{2016MNRAS.456.4156K} to provide a covariance matrix. Given the difference in volume between these simulations, we scale the MD-PATCHY covariance matrix to match the expected errors from the {\sc Aemulus} simulation measurements, applying linear-theory based volume scaling. In this section, we describe both simulation sets further.

\subsection{{\sc Aemulus} $w\mathrm{CDM}$ simulations}

The $w\mathrm{CDM}$ simulations used during this work are from the {\sc Aemulus} project \citep{2018arXiv180405865D}. We use a suite of 40 halo catalogues each with a different cosmology, where $(\Omega_{b}h^{2}, \Omega_{c}h^{2}, w_{0}, n_{s}, \log 10^{10}A_{s}, H_{0}, N_{\mathrm{eff}})$ are varied. The simulations are built on GADGET-2 N-body simulations with $1400^{3}$ particles, periodic box length of $L = 1.05h^{-1}\mathrm{Gpc}$ and a mass resolution of $3.51\times10^{10}(\Omega_{m}/0.3)h^{-1}\mathrm{M}_{\odot}$. For this work we focus on snapshots taken at a redshift $z = 0.55$. 

\begin{figure*}
    \centering
	\includegraphics[width=2\columnwidth]{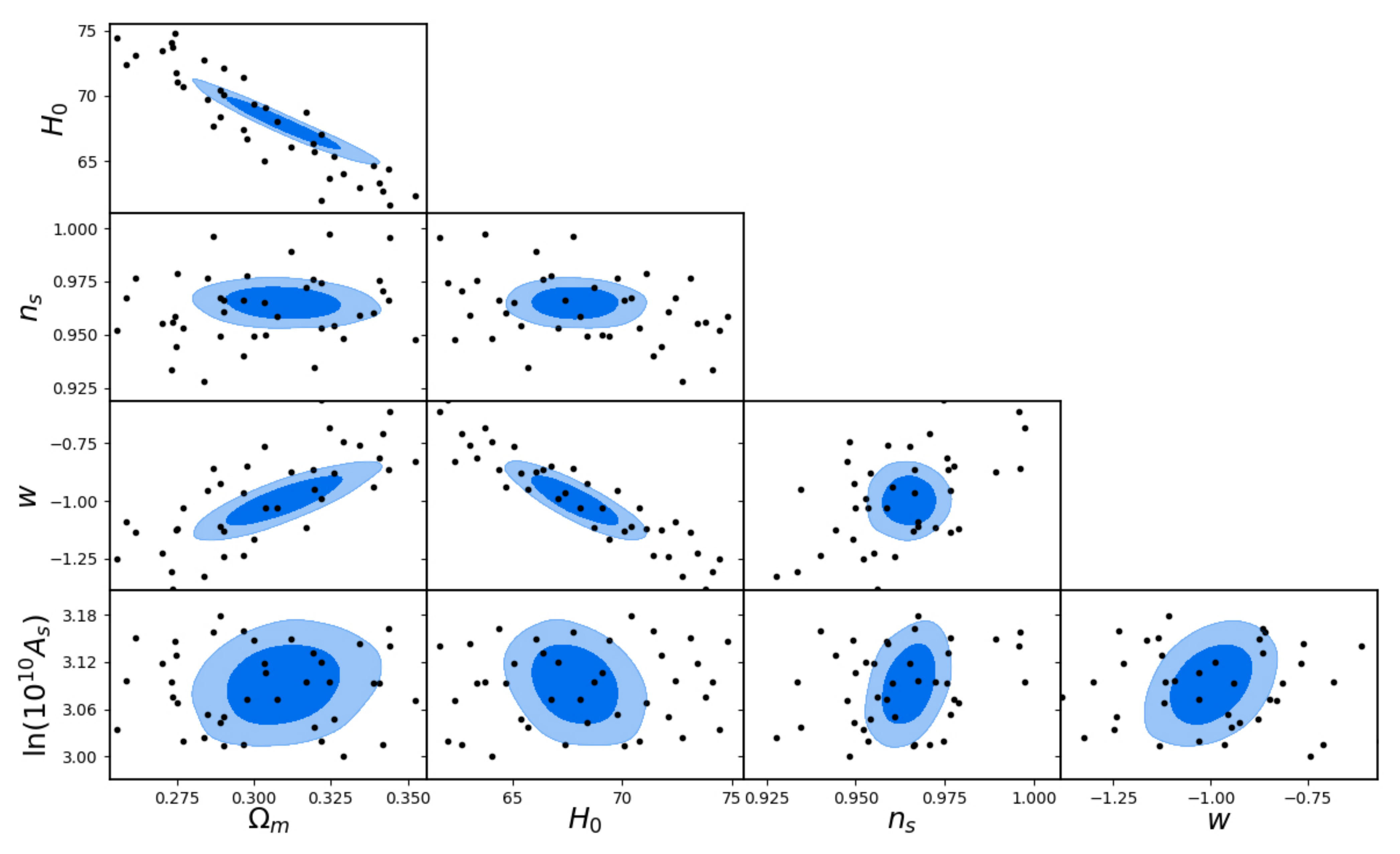}
    \caption{Comparison of the hypercube sampling in cosmological parameters against the likelihood contours of $w\mathrm{CDM}$ with the latest Planck 2018 + BOSS consensus \citep{2017MNRAS.470.2617A, 2018arXiv180706209P}.}
    \label{fig:aem_sample}
\end{figure*}

The cosmologies of these simulations were sampled using a Latin hypercube method \citep{2009ApJ...705..156H} from the joint likelihoods of Planck 2013 and WMAP9 within $4\sigma$ confidence intervals. This allows our tests to effectively sample trends of BAO peak systematics across the 7-dimensional hypercube (can be seen in Figure 3 of \citealt{2018arXiv180405865D}). A comparison between five of the cosmological parameters against a Planck 2018 + BAO consensus \citep{2017MNRAS.470.2617A, 2018arXiv180706209P} is given in Figure \ref{fig:aem_sample}.

The catalogues have been generated by defining dark matter haloes as spherical structures with overdensities 200 times the background density. The halos are located using the {\sc Rockstar} spherical overdensity halo
finder \citep{2013ApJ...762..109B} selected to have typical radii of $\sim 0.5-2h^{-1}\mathrm{Mpc}$. In \cite{2018arXiv180405865D} convergence tests are run to validate the simulations for galaxy clustering studies. Comparisons to training simulations using the HALOFIT algorithm \citep{2003MNRAS.341.1311S, 2012ApJ...761..152T} show agreement to better than $1\%$ in mean deviation up to $k < 0.3h\mathrm{Mpc}^{-1}$.

\subsection{MD-PATCHY halo catalogues}
The 4096 MD-PATCHY catalogues which we use to provide a covariance matrix were originally created in the process of generating MD-PATCHY mocks for analysis of the CMASS BOSS survey \citep{2016MNRAS.456.4156K}.

The catalogues all have the same input power spectrum, but different initial conditions generated by Augmented Lagrangian Perturbation Theory \citep{2014MNRAS.439L..21K} encoded in the PATCHY code. These are then calibrated against the BigMultiDark simulation \citep{2016MNRAS.457.4340K} which were performed using GADGET-2 \citep{2005Natur.435..629S}. The simulation boxes contained $3840^3$ particles in $(2500h^{-1}\mathrm{Mpc})$ with a $\Lambda \mathrm{CDM}$ Planck cosmology, $\Omega_{m}=0.307115$, $\Omega_{b}=0.048206$, $\sigma_{8}=0.8288$, $n_{s}=0.9611$ and $h = 0.6777$. Haloes are then defined based upon the Bound Density Maximum halo finder \citep{1997astro.ph.12217K}. These halo snapshots are chosen to closely match the {\sc Aemulus} simulations in redshift $z = 0.5328$, however, the difference in volume and number density will need to be accounted for through linear covariance matrix scaling.

\section{Methodology}  \label{sec:method}

Our work in this paper focuses on the Fourier space analysis, where we measure BAO in moments of the power spectrum. Although our methodology follows that of many recent papers \citep{2011MNRAS.415.2892B, 2015MNRAS.449..835R, 2016MNRAS.460.4210G, 2017MNRAS.464.3409B}, for completeness we give a brief overview in this section.

\subsection{Power Spectrum Multipoles}

We use an estimator for the power-spectrum multipoles constructed from the weighted galaxy density field as described in \citet{1994ApJ...426...23F},
\begin{equation}
F(\mathbf{r}) = \frac{1}{\sqrt{I}}[n(\mathbf{r})-\bar{n}(\mathbf{r})]\,,
\end{equation}
where $n(\mathbf{r})$ is the observed number density of haloes and $\bar{n}(\mathbf{r})$ is the expected density, which we can easily calculate for the simulations given the number of haloes and the number of grid cells. As the density is constant, we do not apply any weights. $I$ normalises the amplitude of observed power $I = \int d\mathbf{r} \bar{n}^2(\mathbf{r})$. $F(\mathbf{r})$ was constructed on a Cartesian grid, by distributing haloes and randoms using a Cloud-in-cell (CIC) grid assignment scheme \citep{HEBOOK}. An interlacing technique was used to reduce the aliasing effect when calculating Fourier transforms \citep{2016MNRAS.460.3624S}.

The statistics we fit are the power spectrum multipoles of $F(\mathbf{r})$,
\begin{dmath}  \label{eq:estimators}
    \begin{split}
        \hat{P}_{\ell}(k) &= \frac{2\ell+1}{I}\int\frac{d\Omega_{k}}{4\pi}\Bigg[\int d\mathbf{r}_{1}\int d\mathbf{r}_{2} F(\mathbf{r}_{1})F(\mathbf{r}_{2})e^{ik\cdot(\mathbf{r}_{1}-\mathbf{r}_{2})}\times\\
        &\;\;\;\;\mathcal{L}_{\ell}(\hat{\mathbf{k}}\cdot\hat{\mathbf{r}}_{h})-P_{\ell}^{\mathrm{noise}}(k)\Bigg]\,,
    \end{split}
\end{dmath}
where $\ell$ defines the order of multipole taken with respect to the line-of-sight. We make a plane--parallel assumption and take the z-axis of each simulation as the line-of-sight direction both when moving the simulation to redshift-space and when calculating the multipoles. We fit the monopole and quadrupole moments ($\ell = 0, 2$), ignoring the hexadecapole, which contains a low level of BAO information, following the current standard analysis \citep{2017MNRAS.464.3409B}. $\Omega_{k}$ is the solid angle in $k$-space and $\mathcal{L}_{\ell}(\hat{\mathbf{k}}\cdot\hat{\mathbf{r}}_{h})$ is the Legendre polynomial taking the cosine angle to the line-of-sight (LOS). $P_{\ell}^{\mathrm{noise}}(k)$ is the shot noise term for the power spectrum which can be calculated
\begin{equation}
P_{\ell}^{\mathrm{noise}}(k) = (1+\alpha)\int d\mathbf{r} \bar{n}(r)w^{2}(r)\mathcal{L}_{l}(\hat{\mathbf{k}}\cdot\hat{\mathbf{r}_{h}})\,.
\end{equation}

Power spectra were generated using the publicly available nbodykit package\footnote{http://nbodykit.readthedocs.io/en/latest/} \citep{2017arXiv171205834H}. The formalism used to calculate Eq.~\ref{eq:estimators} using fast Fourier transforms (FFTs) is described in \citet{2017JCAP...10..009H}, and builds upon the ideas of \cite{2015MNRAS.453L..11B, 2015PhRvD..92h3532S} and \cite{2015arXiv151004809S}.

\subsection{Covariance Matrix}

The covariance matrix of the $z = 0.55$ {\sc Aemulus} halo catalogues was generated using 4096 MD-PATCHY catalogues at a similar snapshot redshift. The matrix was calculated taking into account the auto and cross-correlation elements of the monopole and quadrupole,
\begin{equation}
    \textbf{C}_{ij} = \sum\limits_{n=1}^{N}\frac{(P_{\ell, n}(k_{i})-\overline{P}_{\ell}(k_{i}))(P_{\ell', n}(k_{j})-\overline{P}_{\ell'}(k_{j}))}{N-1}\,,
    \label{eq:equ5b}
\end{equation}
where the summation runs over $N$ mock realisations. $P_{\ell, n}(k_{i})$ is the $i^{\textrm{th}}$ separation bin in $k$-space of the $n^{\textrm{th}}$ mock power spectrum in the $\ell^{\textrm{th}}$ multipole and $\overline{P}_{\ell}(k_{i})$ is the average in this bin and multipole.

\begin{figure}
    \centering
	\includegraphics[width=\columnwidth]{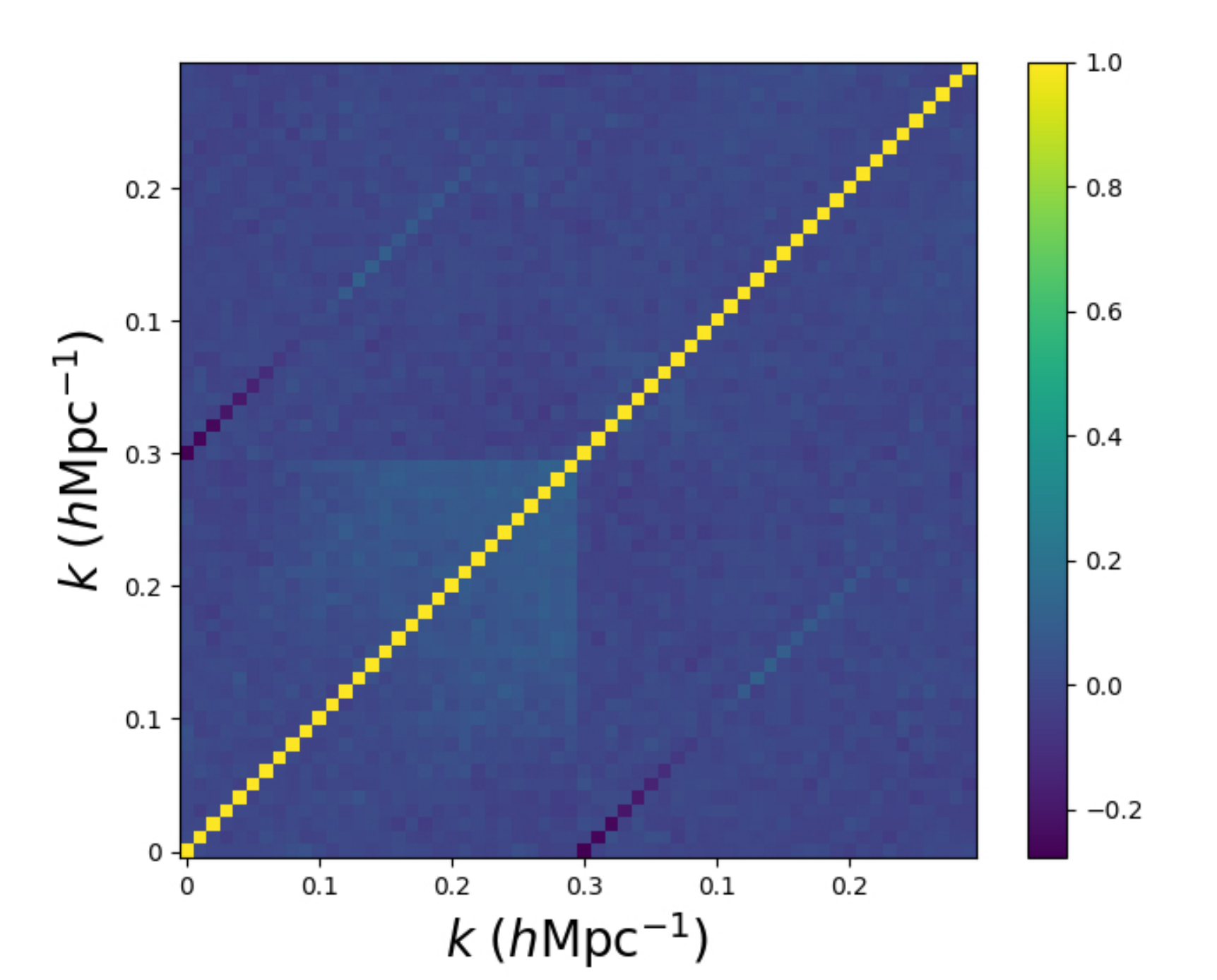}
    \caption{The correlation matrix built from the 4096 Patchy halo catalogues. The bottom left quadrant shows the correlation matrix for the monopole, upper right for the quadrupole and others show the cross-correlation between multipoles. The corresponding covariance matrix is scaled to allow for use with {\sc Aemulus} simulations.}
    \label{fig:figure1}
\end{figure}

There are a number of differences between the MD-PATCHY and {\sc Aemulus} halo catalogues which will affect the covariance matrix, including the number density, volume, and underlying cosmology. The volume and density differences can be incorporated through scaling of the covariance matrix by the ratio of the MD-PATCHY effective volume ($L_{\mathrm{PATCHY}}=2500h^{-1}\mathrm{Mpc}$) to the {\sc Aemulus} effective volume ($L_{w\mathrm{CDM}}=1050h^{-1}\mathrm{Mpc}$). Here, the high density of halos means that the shot noise has a negligible effect on the effective volume. We correct for a broad change in the amplitude of the power spectrum between MD-PATCHY and the average over all 40 {\sc Aemulus} simulations, by scaling the covariance matrix by the ratio of the power spectrum monopole amplitude on linear scales $(k = 0.1h\mathrm{Mpc}^{-1})$. For the {\sc Aemulus} simulations, we calculate the mean power spectrum over all values of the sampled cosmological parameters.
\begin{equation}
    \textbf{C}_{w\mathrm{CDM}}=\left(\frac{L_{\mathrm{PATCHY}}}{L_{w\mathrm{CDM}}}\right)^{3}\left(\frac{P_{0, \mathrm{Aemulus}}}{P_{0, \mathrm{PATCHY}}}\right)\textbf{C}_{\mathrm{PATCHY}}\,.
    \label{eq:equ5b}
\end{equation}
The resulting correlation matrix is shown in Fig.~\ref{fig:figure1}.   

\subsection{Density Field Reconstruction}

The formalism used for density field reconstruction follows from \cite{2014MNRAS.445.3152B, 2015MNRAS.453..456B}. In a Lagrangian framework, the Eulerian position of a particle is given by
\begin{equation}
\textbf{\textrm{x}}(\textbf{\textrm{q}}, t) = \textbf{\textrm{q}} + \boldsymbol{\Psi}(\textbf{\textrm{q}}, t)\,,
\label{eq:equ15}
\end{equation}
where the \textbf{$\textrm{q}$} is the Lagrangian position and $\boldsymbol{\Psi}$ is the displacement vector field. Implementing first order Lagrangian Perturbation Theory (LPT) the standard Zel'dovich approximation \citep{1970A&A.....5...84Z} can be obtained
\begin{equation}
\boldsymbol{\Psi}_{(1)}(\textbf{\textrm{k}})=-\frac{i\textbf{\textrm{k}}}{k^{2}}\delta_{(1)}(\textbf{\textrm{k}}),
\label{eq:equ16}
\end{equation}
which relates the Fourier transform of the overdensity field to the displacement field in $\textbf{\textrm{k}}$-space. To linear order galaxies trace the matter density field as $\delta_{g}=b\delta_{m}$ where $b$ is the bias. To obtain the displacement field $\Psi$ we actually have to solve the differential equation including redshift space distortions (RSDs)
\begin{equation}
\nabla\cdot\boldsymbol{\Psi}+\frac{f}{b}\nabla\cdot(\boldsymbol{\Psi}\cdot\boldsymbol{\hat{r}})\boldsymbol{\hat{r}} = -\frac{\delta_{g}}{b}.
\label{eq:equ17}
\end{equation}
On linear scales RSD enhances the clustering along the line-of-sight, dependent on the amplitude of $f = d\ln(D(a))/d\ln(a)$ the growth rate, $D(a)$ the growth function, $a$ the scale factor and $\sigma_{8}$ describes the amplitude of the density perturbations within spheres of scale $8\,h^{-1}$Mpc.

Equation~\ref{eq:equ17} can be solved as in \cite{2012MNRAS.427.2132P} using a finite difference approximation to compute the gradients. This sets up a grid in configuration space through which the potential can be described as a linear system of equations. This methodology was chosen because although $\boldsymbol{\Psi}$ is irrotational, the term $(\boldsymbol{\Psi}\cdot\boldsymbol{\hat{r}})\boldsymbol{\hat{r}}$ is not, hence one cannot locate the solution directly with Fourier methods. However \cite{2015MNRAS.453..456B} showed that by making the approximation that $(\boldsymbol{\Psi}\cdot\boldsymbol{\hat{r}})\boldsymbol{\hat{r}}$ is irrotational and iterating after correcting, one can efficiently obtain the correct solution using FFTs (with $\textrm{IFFT}$ referring to the inverse fast fourier transform) with $\beta = f/b$
\begin{equation}
\boldsymbol{\Psi}=\textrm{IFFT}\left[-\frac{i\textbf{\textrm{k}}\delta(k)}{k^{2}b}\right]-\frac{\beta}{1+\beta}\left(\textrm{IFFT}\left[-\frac{i\textbf{\textrm{k}}\delta(k)}{k^{2}b}\right]\cdot\hat{\textbf{r}}\right)\hat{\textbf{r}}.
\label{eq:equ18}
\end{equation}
The displacement field calculated from this form of the algorithm has been shown to agree with the finite difference approach and causes negligible differences between post-reconstruction 2-point statistics \citep{2014MNRAS.445.3152B}.

To remove RSD we modify the displacement vector as $\Psi^{\mathrm{final}}=\Psi+\Psi_{\mathrm{RSD}}$ \citep{1987MNRAS.227....1K, 2012MNRAS.427.2132P} where
\begin{equation}
\Psi_{\mathrm{RSD}} = -f(\mathbf{\Psi}\cdot\mathbf{\hat{r}})\mathbf{\hat{r}},
\label{eq:equ176}
\end{equation}
using the already calculated displacement field along the line-of-sight. This retrieval of the real-space post-reconstruction density field results in the reduction of amplitude in the power spectrum.

\subsubsection{Dependence on the Fiducial Cosmology}
The fiducial cosmology enters the density field reconstruction procedure at three points in the process.
\begin{enumerate}
    \item The measured redshift has to be converted to a distance to allow a galaxy to be placed on the Cartesian grid. In our case of a halo box (which starts from Cartesian coordinates) this transformation is emulated by an anisotropic scaling of the coordinate system. Under the plane-parallel assumption, taking the $z$-axis as the effective line-of-sight, we scale the $x, y$ plane by the ratio $D_{A, \mathrm{fid}}/D_{A, \mathrm{true}}$ and scale along the $z$ axis by a ratio $H_{\mathrm{true}}/H_{\mathrm{fid}}$. Here the subscript $\mathrm{true}$ and $\mathrm{fid}$ refer to assuming the true and fiducial cosmologies

    \item One of the input parameters to reconstruction is the linear bias of the galaxies. In our $w\mathrm{CDM}$ halo catalogues we work in redshift space, and in linear theory the pre-reconstruction halo power spectrum $P_{h}(k)$ and matter power spectrum $P_{m}(k)$ are related by
    \begin{equation}
        P_{h}(k,\mu) = \left(b + f\mu^{2}\right)^{2}P_{m}(k)\,,
    \end{equation}
    where $\mu$ is the cosine of the angle of the mode to the line-of-sight. The matter power spectrum is estimated using CLASS \citep{2011arXiv1104.2932L, 2011JCAP...07..034B}, so there is a dependence on the assumed cosmology at this stage. To calculate the bias for each pair of true and assumed cosmology, the halo power spectrum monopole is calculated and fitted to minimise the difference between $P_{0, h}$ and $P_{0, m}$.
    
    \item The other input parameter through which the fiducial cosmology assumption appears is the linear growth rate, $f$. This growth rate also enters in the calculation of linear bias as seen in (ii).
\end{enumerate}
These assumptions have a direct impact on the calculation of the displacement field $\mathbf{\Psi}$, and the RSD signal.

\subsection{Model fitting}

The key parameters measured when fitting BAO are $\alpha_{\parallel}$ and $\alpha_{\bot}$. These parameters are used in standard BAO studies to approximately quantify deviations in scale between the measured power spectrum and the template calculated to match the fiducial cosmology,
\begin{equation}
    \alpha_{\parallel} = \frac{H^{\mathrm{fid}}(z)r_{s}^{\mathrm{fid}}(z_{d})}{H(z)r_{s}(z_{d})}\,,
\end{equation}
\begin{equation}
    \alpha_{\bot} = \frac{D_{A}(z)r_{s}^{\mathrm{fid}}(z_{d})}{D_{A}^{\mathrm{fid}}(z)r_{s}(z_{d})}\,.
\end{equation}
To measure these we perform a joint fit of the monopole and quadrupole, where we scale the wave-numbers of the model by $k'_{\parallel} = k_{\parallel}/\alpha_{\parallel}$, and $k'_{\bot} = k_{\bot}/\alpha_{\bot}$ to match the data. For the anisotropic fitting, we initially need a template for the 2D power spectrum, therefore, it is useful to define,
\begin{equation}
    k' = \frac{k}{\alpha_{\bot}}\left[1+\mu^{2}\left(\frac{\alpha_{\bot}^{2}}{\alpha_{\parallel}^{2}}-1\right)\right] ^{1/2}\,,
\end{equation}    
\begin{equation}
    \mu' = \frac{\mu\alpha_{\bot}}{\alpha_{\parallel}}\left[1+\mu^{2}\left(\frac{\alpha_{\bot}^{2}}{\alpha_{\parallel}^{2}}-1\right)\right] ^{-1/2}\,.
\end{equation}    

The template for the anisotropic power spectrum is given by
\begin{dmath}
    P(k, \mu) = P_{\mathrm{sm}}(k, \mu)\times\left[1+(\mathcal{O}_{\mathrm{lin}}(k)-1)
    e^{-[k^{2}\mu^{2}(1+f)^{2}\Sigma_{\bot}^{2}+k^{2}(1-\mu^{2})\Sigma_{\bot}^{2}]/2}
    \right]\,,
\end{dmath}
where $\mathcal{O}_{\mathrm{lin}}(k)$ represents the oscillatory part of the fiducial linear power spectrum which is obtained by fitting $P_{\mathrm{lin}}(k)$ with a \citep{1998ApJ...496..605E} no-wiggle power spectrum with five polynomial terms to get $P_{\mathrm{sm, lin}}(k)$ and then take the ratio, $\mathcal{O}_{\mathrm{lin}}(k) = P_{\mathrm{lin}}(k)/P_{\mathrm{sm, lin}}(k)$. $\Sigma_{\bot}$ is the non-linear damping term across the line-of-sight, in this template we have used the form as in \cite{2018MNRAS.479.1021D} where the term along the line-of-sight $\Sigma_{\parallel} = (1+f)\Sigma_{\bot}$. The smooth anisotropic power spectrum, $P_{\mathrm{sm}}(k, \mu)$ is given by
\begin{dmath}  \label{eq:comovP}
    P_{\mathrm{sm}}(k, \mu) = B^{2}\left(1 + \beta\mu^{2}\left[1-\exp(-(k\Sigma_{\mathrm{smooth}})^{2}/2)\right]\right)^{2}\times P_{\mathrm{sm, lin}}(k)F_{\mathrm{damp}}(k, \mu, \Sigma_{s}),
\end{dmath}
for a post-reconstruction power spectrum. The $B$ parameter is used to marginalise over the power spectrum amplitude and $\Sigma_{\mathrm{smooth}}$ is the length scale used in the smoothing kernel during reconstruction, here $\Sigma_{\mathrm{smooth}} = 15h^{-1}\mathrm{Mpc}$. $F_{\mathrm{damp}}(k, \mu, \Sigma_{s})$ is the damping term due to the non-linear velocity field (Finger-of-God). In this work, we take this term as
\begin{equation}
  F_{\mathrm{damp}}(k, \mu, \Sigma_{s})
  =\left\{
  \begin{array}{@{}ll@{}}
      \left[\left(1+k^{2}\mu^{2}\Sigma_{s}^{2}\right)/2\right]^{-2}\,, & \text{if}\ \Sigma_{s}>0\,, \\
      [1pt]\\
      \left[\left(1+k^{2}\mu^{2}\Sigma_{s}^{2}\right)/2\right]^{2}\,, & \text{if}\ \Sigma_{s}<0\,,
    \end{array}
    \right.
\end{equation}
which is an extension upon the usual definition, which only considers the region where $\Sigma_{s}>0$. This extension allows for differences between the fiducial and true models, where the true model has a BAO feature that is sharper than that in the fiducial model. If this definition is not used, we sometimes see a set of best-fit solutions clustered at $\Sigma_{s}=0$, which can bias the average results. 

From the anisotropic power spectrum, we obtain templates for the monopole and quadrupole. For each, we include 5 polynomial terms which allow for marginalisation over the broadband shape and variations between the cosmological models other than the BAO position
 \begin{equation}
    P_{0}(k) = \frac{1}{2}\int\limits_{-1}^{1}P(k, \mu)d\mu + A_{0}(k)\,,
\end{equation} 
\begin{equation}
    P_{2}(k) = \frac{5}{2}\int\limits_{-1}^{1}P(k, \mu)\mathcal{L}_{2}(\mu)d\mu + A_{2}(k)\,,
\end{equation}
where the polynomial terms are
\begin{equation}
    A_{\ell}^{\mathrm{post-recon}}(k) = \frac{a_{\ell, 1}}{k^{3}} + \frac{a_{\ell, 2}}{k^{2}} + \frac{a_{\ell, 3}}{k} + a_{\ell, 4} + a_{\ell, 5}k^{2}\,.
\end{equation}

Fits between model and data were conducted using the Affine Invariant Markov chain Monte Carlo (MCMC) Ensemble sampler within the \texttt{emcee} package \citep{2013PASP..125..306F}. The MCMC algorithm uses a Gelman-Rubin convergence criteria \citep{gelman1992} between 4 chains, based upon the in-chain and cross-chain variances. Once this metric passes below a defined threshold, in this case $\epsilon = 0.015$, the chain is considered to have reached convergence. The model has 14 free parameters [$B$, $a_{0, 1}$, $a_{0, 2}$, $a_{0, 3}$, $a_{0, 4}$, $a_{0, 5}$, $a_{2, 1}$, $a_{2, 2}$, $a_{2, 3}$, $a_{2, 4}$, $a_{2, 5}$, $\Sigma_{\bot}$, $\alpha_{\parallel}$, $\alpha_{\bot}$, $\Sigma_{s}$], with $\beta$ and $f$ fixed to the values in the fiducial cosmology, and $\Sigma_{\mathrm{smooth}}=15$\,h$^{-1}$Mpc. Both the monopole and quadrupole are simultaneously fit between $0.01h\mathrm{Mpc}^{-1} < k < 0.3h\mathrm{Mpc}^{-1}$ in bins of $\Delta k = 0.01h\mathrm{Mpc}^{-1}$.

To ensure that the inversion of the covariance matrix is unbiased during fitting we apply scaling as described in \citep{2007A&A...464..399H}
\begin{equation}
    \mathbf{C}_{i, j, \mathrm{Hartlap}}^{-1} = \frac{N_{s}-n_{b}-2}{N_{s}-1}C_{i, j}^{-1}\,.
\end{equation}  
Here $n_{b}$ is the number of power spectrum bins $n_{b}=60$ and $N_{s}$ is the number of MD-PATCHY catalogues used $N_{s}=4096$. The correction increases the diagonal variance by $\sim 1.5\%$.

\section{Testing the cosmology assumed for reconstruction}
\label{sec:resRec}

To search for systematic biases that arise from the mismatch of true and assumed cosmology during reconstruction we need to obtain an effective sampling of potential differences between true and assumed cosmologies. The 40 {\sc Aemulus} simulations were sampled from the Planck13 \citep{2014A&A...571A..16P} and WMAP9 \citep{2013ApJS..208...19H} joint likelihoods, and we use these models also for the assumed cosmologies. Therefore we run the reconstruction and BAO fitting pipeline 1600 times using 40 different assumptions of cosmology for each of the 40 simulations. 40 of the 1600, therefore, have the true cosmology as the base assumption.

For the results presented in this section, our aim is to look at the biases coming from the reconstruction process only, and so the data box is re\-scaled back to the true cosmology coordinate frame and fitting is performed assuming the correct comoving power spectrum of Eq.~\ref{eq:comovP}. 

\subsection{The distribution of recovered parameters}

If assuming a different fiducial cosmology when performing reconstruction does not bias results, then we expect to recover $\alpha_{\parallel}=1$ and  $\alpha_{\bot}=1$ on average with some scatter due to sample variance. In order to reduce the sample variance, we study $\Delta \alpha_{\parallel} = \alpha_{\parallel, \mathrm{fid}} - \alpha_{\parallel, \mathrm{true}}$ and $\Delta \alpha_{\bot} = \alpha_{\bot, \mathrm{fid}} - \alpha_{\bot, \mathrm{true}}$ rather than the scaling parameters $\alpha$ themselves. By looking at the distribution of $\Delta \alpha_{\parallel}$ and $\Delta \alpha_{\bot}$ we would expect to see no shift on average over the 1600 combinations as we are likely averaging biases in different directions. The standard deviation of the distribution gives us an idea of the inherent noise on the measured shift from each combination. The overall standard deviation of this distribution will be biased larger as the 40 cosmologies sample the underlying CMB experiment likelihoods using a uniform approach.

\begin{figure}
    \centering
	\includegraphics[width=\columnwidth]{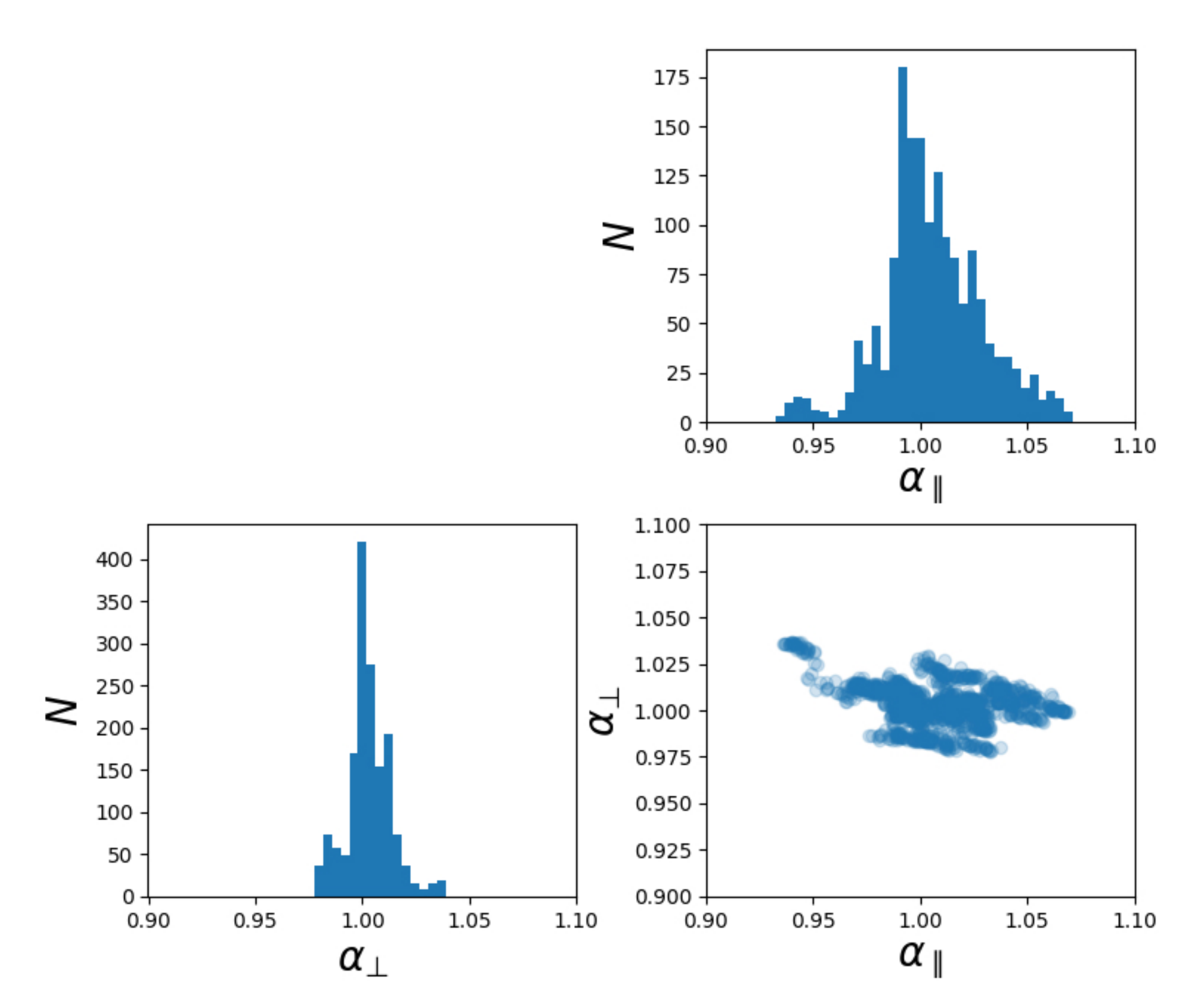}
    \caption{Distributions of measured $\alpha_{\bot}$ and $\alpha_{\parallel}$ for the 1600 combinations of true box and assumed cosmologies used during reconstruction only. The distributions have $\alpha_{\bot}\pm\sigma_{\alpha_{\bot}} = 1.0029\pm0.010$ and $\alpha_{\parallel}\pm\sigma_{\alpha_{\parallel}} = 1.0052\pm0.024$ and scatter around 1 as expected. The clusters seen in the 2D distribution correspond to individual halo boxes and the sample variance in each sample.}
    \label{fig:alp_dist}
\end{figure}

\begin{figure}
    \centering
	\includegraphics[width=\columnwidth]{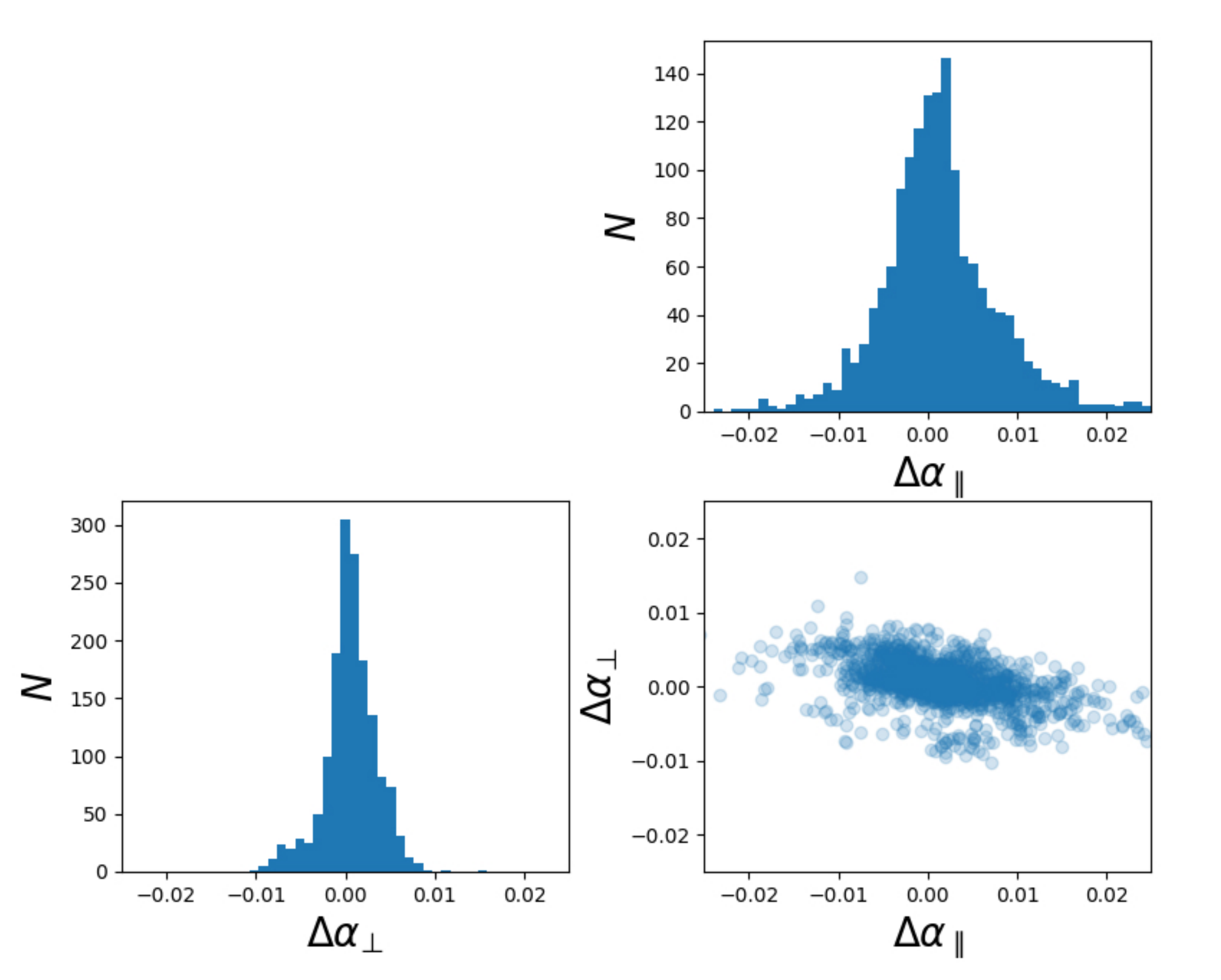}
    \caption{Distributions of measured $\Delta\alpha_{\bot}$ and $\Delta\alpha_{\parallel}$ for the 1600 combinations of true box and assumed cosmologies used during reconstruction only. These distributions have $\Delta\alpha_{\bot}\pm\sigma_{\Delta\alpha_{\bot}} = 0.00060\pm0.0029$ and $\Delta\alpha_{\parallel}\pm\sigma_{\Delta\alpha_{\parallel}} = 0.0013\pm0.0069$ and scatter around 0 as expected.}
    \label{fig:delalp_dist}
\end{figure}

The distribution of $\alpha_{\bot}$ and $\alpha_{\parallel}$ in Figure~\ref{fig:alp_dist}, show an expected scatter around $1$. The variance on the distribution parallel to the line-of-sight is wider because the constraints are intrinsically weaker (using information in only one dimension rather than two). These distributions have $\alpha_{\bot}\pm\sigma_{\alpha_{\bot}} = 1.0029\pm0.010$ and $\alpha_{\parallel}\pm\sigma_{\alpha_{\parallel}} = 1.0052\pm0.024$. The distribution of $\Delta \alpha_{\bot}$ and $\Delta \alpha_{\parallel}$ in Figure~\ref{fig:delalp_dist}, show an expected scatter around $0$. These distributions have $\Delta\alpha_{\bot}\pm\sigma_{\Delta\alpha_{\bot}} = 0.00060\pm0.0029$ and $\Delta\alpha_{\parallel}\pm\sigma_{\Delta\alpha_{\parallel}} = 0.0013\pm0.0069$. The factor of four gain in precision between the scaling parameters $\alpha$ and $\Delta\alpha$ comes from the removal of sample variance in the measurement.

\subsection{Trends as a function of cosmological parameters}  \label{sec:trends-recon}

\begin{figure*}
    \centering
	\includegraphics[width=2.1\columnwidth]{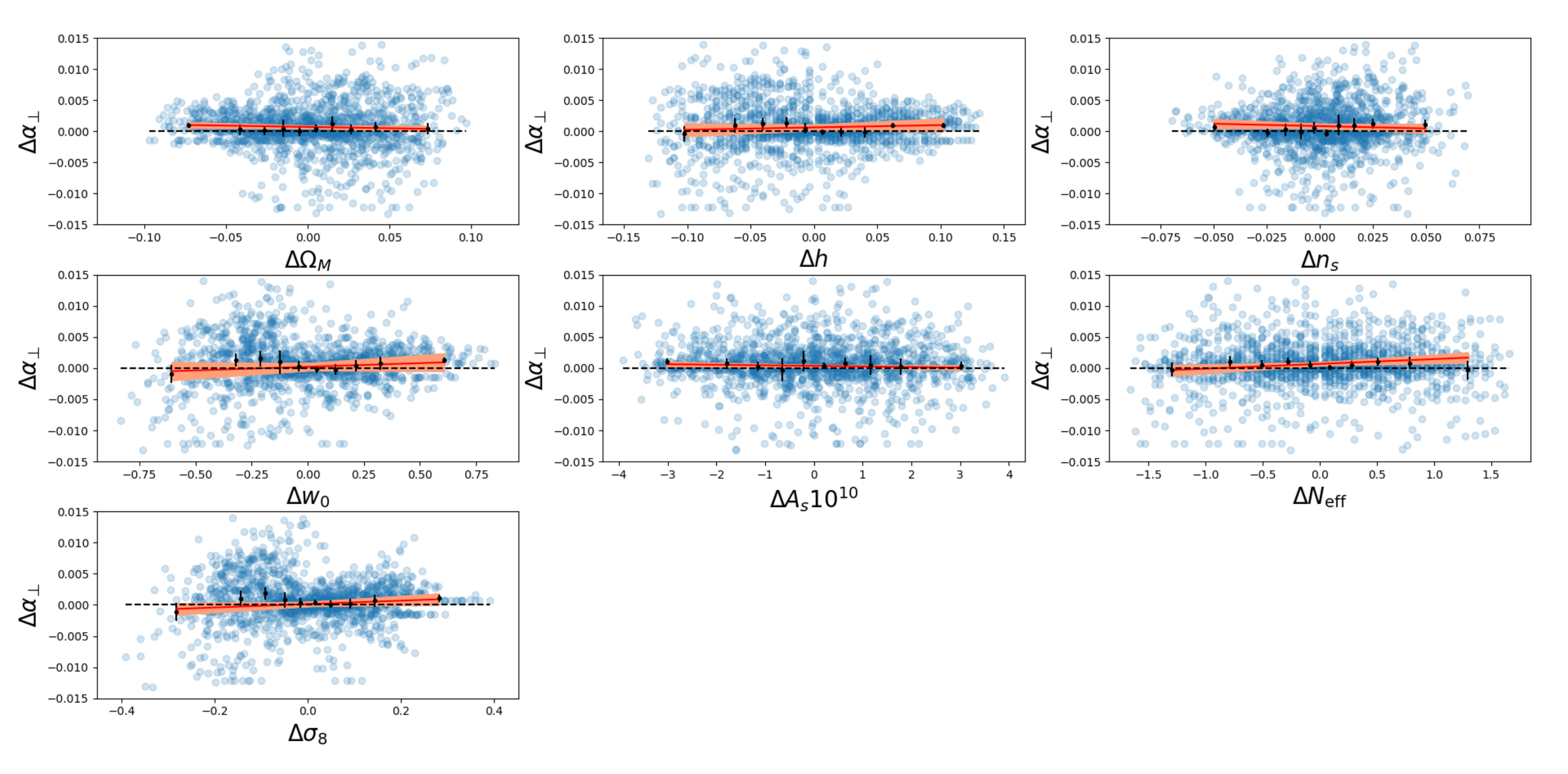}
    \caption{The distribution of $\Delta\alpha_{\bot}$ plotted against the cosmological parameters which have been varied in the {\sc Aemulus} simulations, for tests on reconstruction only. The 1600 scatter points have also been binned in 10 equally sized bins (black data points). The errorbars on these points correspond to the square root of the diagonal elements of the jack-knife re-sample generated covariance matrix. The solid lines with errors show a fit for a linear trend to the black data points. Comparison between the best fit linear trend and a zero-bias flat model show that there is mild evidence for a non-zero systematic bias at the $\Delta\alpha_{\bot}\sim 0.001$ level only for variations in $\Omega_{m}$ and $h$.}
    \label{fig:trend}
\end{figure*}

\begin{figure*}
    \centering
	\includegraphics[width=2.1\columnwidth]{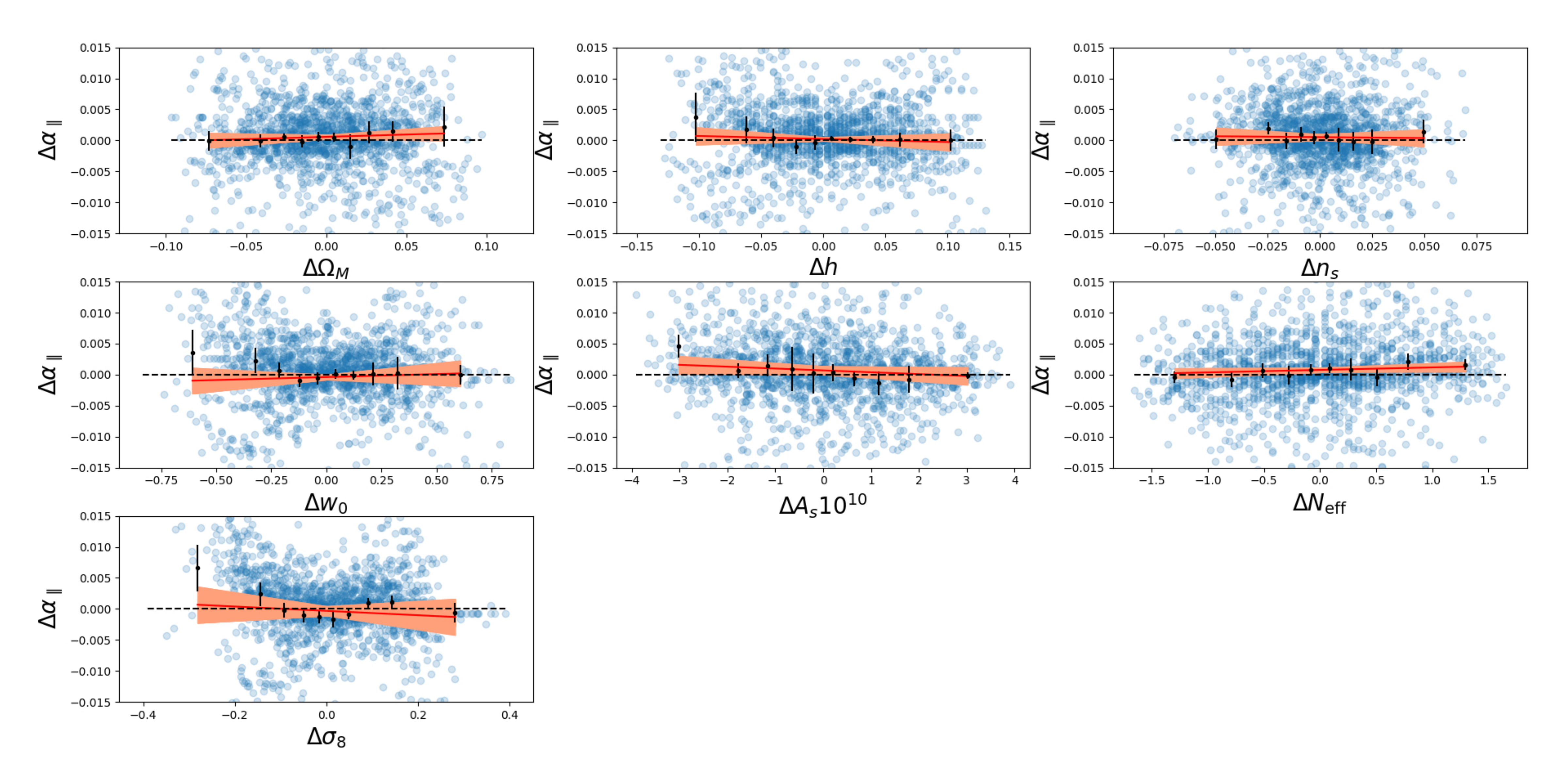}
    \caption{The distribution of $\Delta\alpha_{\parallel}$ against the cosmological parameters which have been varied in the {\sc Aemulus} simulations, for tests on reconstruction only. The 1600 scatter points have also been binned in 10 equally sized bins (black data points). The errorbars on these points correspond to the square root of the diagonal elements of the jack-knife re-sample generated covariance matrix. Comparison between the best fit linear trend and a zero-bias flat model show no evidence for deviations for a unbiased measurement of $\alpha_{\parallel}$.}
    \label{fig:trend2}
\end{figure*}

In order to understand whether systematic biases through assuming an incorrect fiducial cosmology arise from changes in specific cosmological parameters during reconstruction, it is useful to look at the relationship of $\Delta \alpha_{\parallel}$ and $\Delta \alpha_{\bot}$ with respect to $\Delta X$ where $X$ are the cosmological parameters varied in the simulation sampling $\Omega_{m}, w_{0}, n_{s}, \log 10^{10}A_{s}, H_{0}, N_{\mathrm{eff}}$ and $\sigma_{8}$. In Figures \ref{fig:trend} and \ref{fig:trend2}, the distributions of $\Delta \alpha_{\bot}$ and $\Delta \alpha_{\parallel}$ are given respectively. These plots show both the raw scatter and data binned into 10 bins with equal numbers of combinations (160 each), allowing for a similar level of statistical error. There is a strong correlation between bins due to 40 of the 1600 combinations having the same underlying simulation box, and a weaker correlation due to the same fiducial cosmology assumed. To fit a linear trend to these bins and test for non-zero deviations, the correlation is accounted for through the use of a covariance matrix built from jackknife resampling \citep{2009MNRAS.396...19N}. The bins are recalculated with each box being removed from the full combination sample and the covariance matrix calculated as
\begin{equation}
    \textbf{C}_{ij} = \frac{N-1}{N}\sum\limits_{n=1}^{N}(x_{i, n}-\overline{x_{i}})(x_{j, n}-\overline{x_{j}})\,,
    \label{eq:equ5b}
\end{equation}
where the sum runs over the $N=40$ different underlying boxes being removed from the full combination sample and the pre-factor $(N-1)/N$ accounting for the remaining high correlation. The scatter plots in each cosmological parameter are shown in Figure \ref{fig:trend} and Figure \ref{fig:trend2}, these include the binned trends and linear fits. These fits show that even for large deviations in fiducial cosmological parameters during reconstruction, one would expect deviations of $\lessapprox 0.1\%$ in $\Delta\alpha_{\bot}$ and $\Delta\alpha_{\parallel}$. These shifts, however, are extreme considering current limits by Planck 2018 \citep{2018arXiv180706209P} and for more realistic discrepancies between the underlying cosmology and assumed fiducial reconstruction parameters the systematics are negligible.

Using the Akaike Information Criterion (AIC) \citep{1100705}
\begin{equation}
    AIC = \chi^{2}+2k\,,
    \label{eq:equ5b}
\end{equation}
and Bayesian Information Criterion (BIC) \citep{schwarz1978}
\begin{equation}
    BIC = \chi^{2}+k\ln{(n)}\,,
    \label{eq:equ5b}
\end{equation}
model comparisons can be made between the best fit linear trend and a flat model at $\Delta\alpha_{\parallel/\bot}=0$. In these equations $k$ is the number of model parameters ($k=2$ for linear model and $k=0$ for flat model) and $n$ is the number of data points, $n=10$. The model comparison is made by calculating the $\Delta \mathrm{AIC}$ and $\Delta \mathrm{BIC}$ between flat and linear models, these are given for each cosmological parameter for $\Delta\alpha_{\parallel}$ and $\Delta\alpha_{\bot}$ in Table~\ref{tab:1}. The only cosmological parameters that show mild evidence ($\Delta\mathrm{AIC}$,$\Delta\mathrm{BIC}>5$ and $\sqrt{\Delta\chi^{2}}>3$) for a deviation from $\Delta\alpha_{\parallel/\bot} = 0$ are $\Delta\Omega_{m}$ and $\Delta h$ in the case of $\Delta\alpha_{\bot}$.

\begin{table}
 \centering
 \caption{Tabulated results of the model comparison tests between the best fit linear trend and a flat no bias line at $\Delta\alpha = 0$. For each of the 1600 combinations measurements of $\Delta\alpha_{\bot}$ and $\Delta\alpha_{\parallel}$, in the case where we are testing reconstruction only, we give the AIC, BIC and $\sqrt{\Delta\chi^{2}}$ as complimentary comparison indicators. Minor evidence is shown in some cases of variations in cosmological parameters for $\Delta\alpha_{\bot}$, although these trends are in general driven by extreme shifts in underlying model.}
 \label{tab:1}
 \begin{tabular}{cccccccc}
  \hline
   & & $\alpha_{\bot}$ & & \vline & & $\alpha_{\parallel}$ &\\
  \hline
   & $\Delta \mathrm{AIC}$ & $\Delta \mathrm{BIC}$ & $\sqrt{\Delta\chi^{2}}$ & \vline & $\Delta \mathrm{AIC}$ & $\Delta \mathrm{BIC}$ & $\sqrt{\Delta\chi^{2}}$\\ 
  \hline
  $\Delta \Omega_{m}$ & 8.9 & 8.3 & 3.59 & & -2.7 & -3.3 & 1.16\\
  $\Delta h$ & 6.6 & 6.0 & 3.25 & & -3.5 & -4.1 & 0.73\\
  $\Delta n_{s}$ & 4.6 & 4.0 & 2.93 & & -2.7 & -3.3 & 1.15\\
  $\Delta w_{0}$ & 0.9 & 0.3 & 2.21 & & -2.5 & -3.1 & 1.21\\
  $\Delta A_{s}10^{10}$ & 0.9 & 0.3 & 2.21 & & -2.2 & -2.8 & 1.33\\
  $\Delta N_{\mathrm{eff}}$ & 4.1 & 3.5 & 2.84 & & -0.4 & -1.0 & 1.90\\
  $\Delta \sigma_{8}$ & -0.5 & -1.1 & 1.90 & & -2.0 & -2.6 & 1.41\\
  \hline
 \end{tabular}
\end{table}

\begin{figure*}
    \centering
	\includegraphics[width=2\columnwidth]{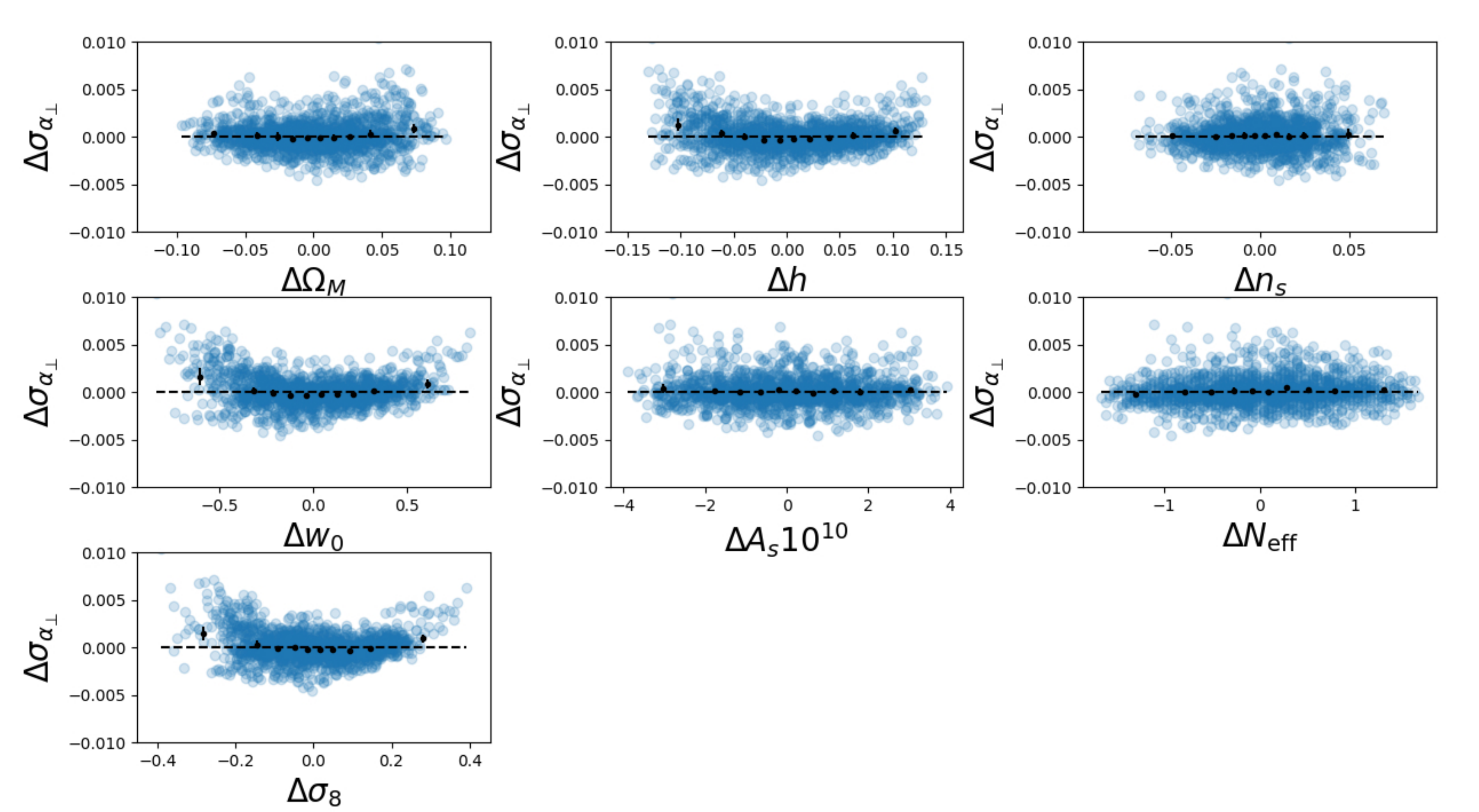}
    \caption{The distribution of $\Delta\sigma_{\alpha_{\bot}}$ against the cosmological parameters which have been varied in the {\sc Aemulus} simulations, for tests on reconstruction only. The 1600 scatter points have also been binned in 10 equally sized bins (black data points). The errorbars on these points correspond to the square root of the diagonal elements of the jack-knife re-sample generated covariance matrix. There can be seen significant deviations, in the case of large differences between cosmologies, from a zero-bias trend. Suggested evidence for an increase by up to $\Delta\sigma_{\alpha_{\bot}} = +0.001$ when incorrect cosmology leads to inefficient reconstruction.}
    \label{fig:trend3}
\end{figure*}

\begin{figure*}
    \centering
	\includegraphics[width=2\columnwidth]{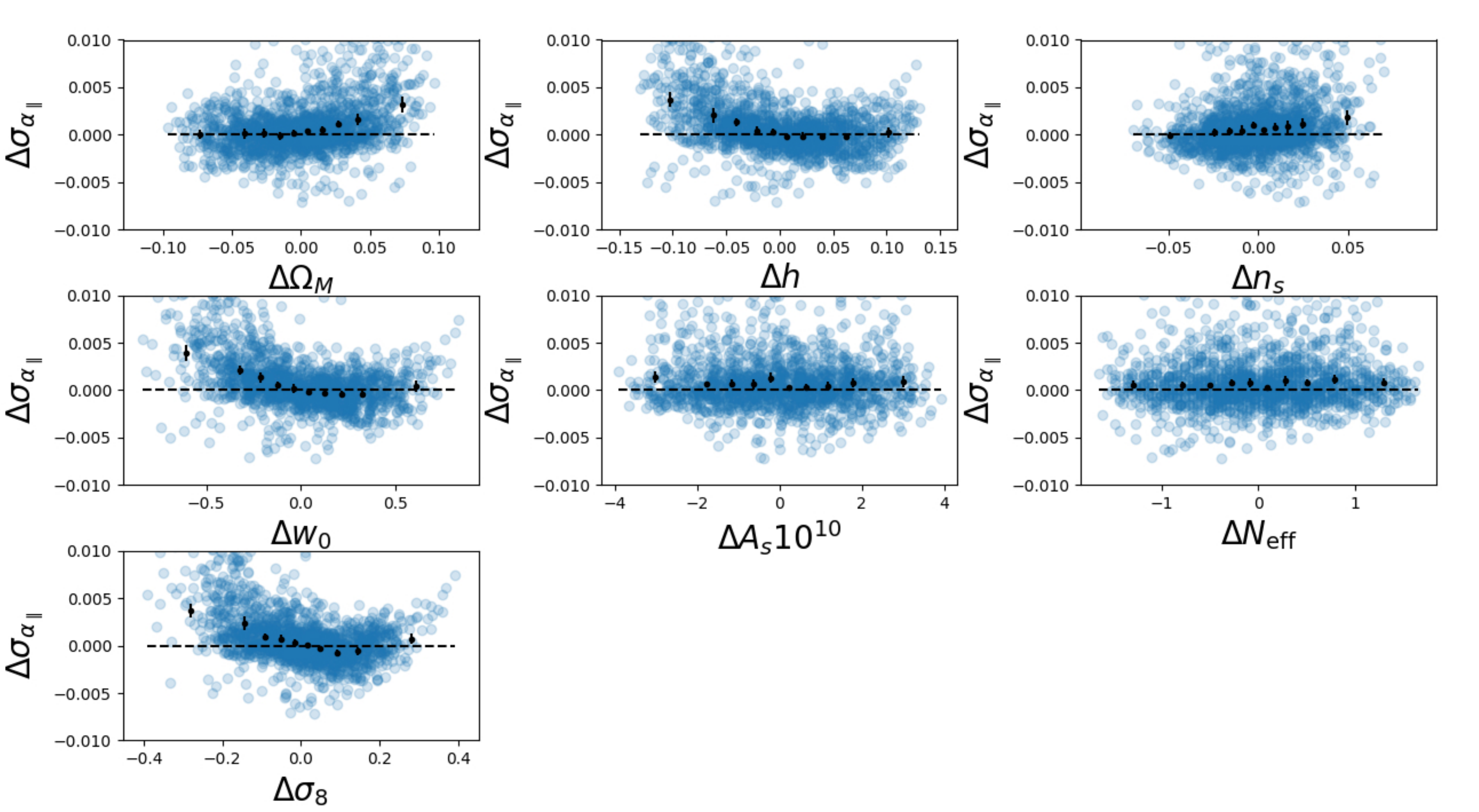}
    \caption{The distribution of $\Delta\sigma_{\alpha_{\parallel}}$ against the cosmological parameters which have been varied in the {\sc Aemulus} simulations, for tests on reconstruction only. The 1600 scatter points have also been binned in 10 equally sized bins (black data points). The errorbars on these points correspond to the square root of the diagonal elements of the jack-knife re-sample generated covariance matrix. There can be seen significant deviations, in the case of large differences between cosmologies, from a zero-bias trend. Suggested evidence for an increase by up to $\Delta\sigma_{\alpha_{\parallel}} = +0.002$ when incorrect cosmology leads to inefficient reconstruction.}
    \label{fig:trend4}
\end{figure*}

From the likelihoods of each MCMC run, the error on $\alpha$, $\sigma_{\alpha}$, can be measured. As well as looking at systematic biases from the incorrect fiducial cosmology in reconstruction for $\alpha$, we can also look for changes in $\sigma_{\alpha}$, indicating whether the wrong cosmology leads to a reduced precision in the BAO scale measurement. In Figure~\ref{fig:trend3} and Figure~\ref{fig:trend4}, the trend in $\Delta\sigma_{\alpha_{\bot}}$ and $\Delta\sigma_{\alpha_{\parallel}}$ are shown. For both $\sigma_{\alpha_{\bot}}$ and $\sigma_{\alpha_{\parallel}}$ there is evidence for an increased error in the case of the fiducial $h$, $w_{0}$ and $\sigma_{8}$ being less than the true cosmological value.

\section{Results from different assumed cosmologies in reconstruction and fitting}
\label{sec:resFit}

Section~\ref{sec:resRec} showed that for density field reconstruction alone, incorrect assumptions of fiducial cosmology contribute negligible systematic biases on $\Delta\alpha_{\parallel}$ and $\Delta\alpha_{\bot}$. We also found weak evidence for increased errors on $\alpha$ in the case of the fiducial $h$, $w_{0}$ and $\sigma_{8}$ being lower than the truth. However, for realistic surveys, the assumption of incorrect fiducial cosmology would permeate further through the analysis pipeline than just the reconstruction step. In particular, the power spectrum is measured within a scaled coordinate reference frame and also the fitting template uses the fiducial cosmology.

To test whether the assumption of an incorrect fiducial cosmology contributes at these later analysis stages, we emulated these steps using our samples. As undertaken in the reconstruction step considered alone in the previous section, we now rescale the coordinate system used when analysing the data to mimic the AP effect \citep{1979Natur.281..358A} of an incorrect fiducial cosmology. This also changes the underlying sample variance of the field, meaning that our method of removing some of the contributions to the statistical error by examining the differences in measurements between assuming the true or incorrect fiducial models by differencing would not work as effectively. This secondary effect of measuring the power spectrum in a rescaled space due to the AP effect does not contribute strongly on BAO scales and so we opt to scale the field back to the true frame following reconstruction in order to maximise the constraining power of the BAO measurements that we can make from the {\sc Aemulus} simulations.

Once the power spectrum has been measured in the true coordinates, the fitting is then performed assuming the fiducial cosmology. As in the previous section, $\Delta\alpha_{\bot}$ and $\Delta\alpha_{\parallel}$ can be measured as a difference between the pipeline using the true and assumed cosmology. In order for this comparison to be made the measured $\alpha$ needs to be scaled by the expected value so that the scaled measurements have the same expected value. We also take into account that units of $\mathrm{Mpc}^{-1}$ also need to be used carefully to allow a matched definition of recovered parameters. Finally, $\alpha$ also needs to scale by a ratio of the sound horizon in the different cosmologies,
\begin{equation}
    \alpha_{\bot/\parallel}^{(t)} = \frac{r_{d}^{(t)}}{r_{d}^{(a)}}\alpha_{\bot/\parallel}^{(a)}\,.
    \label{eq:equ5b}
\end{equation}

\begin{figure}
    \centering
	\includegraphics[width=\columnwidth]{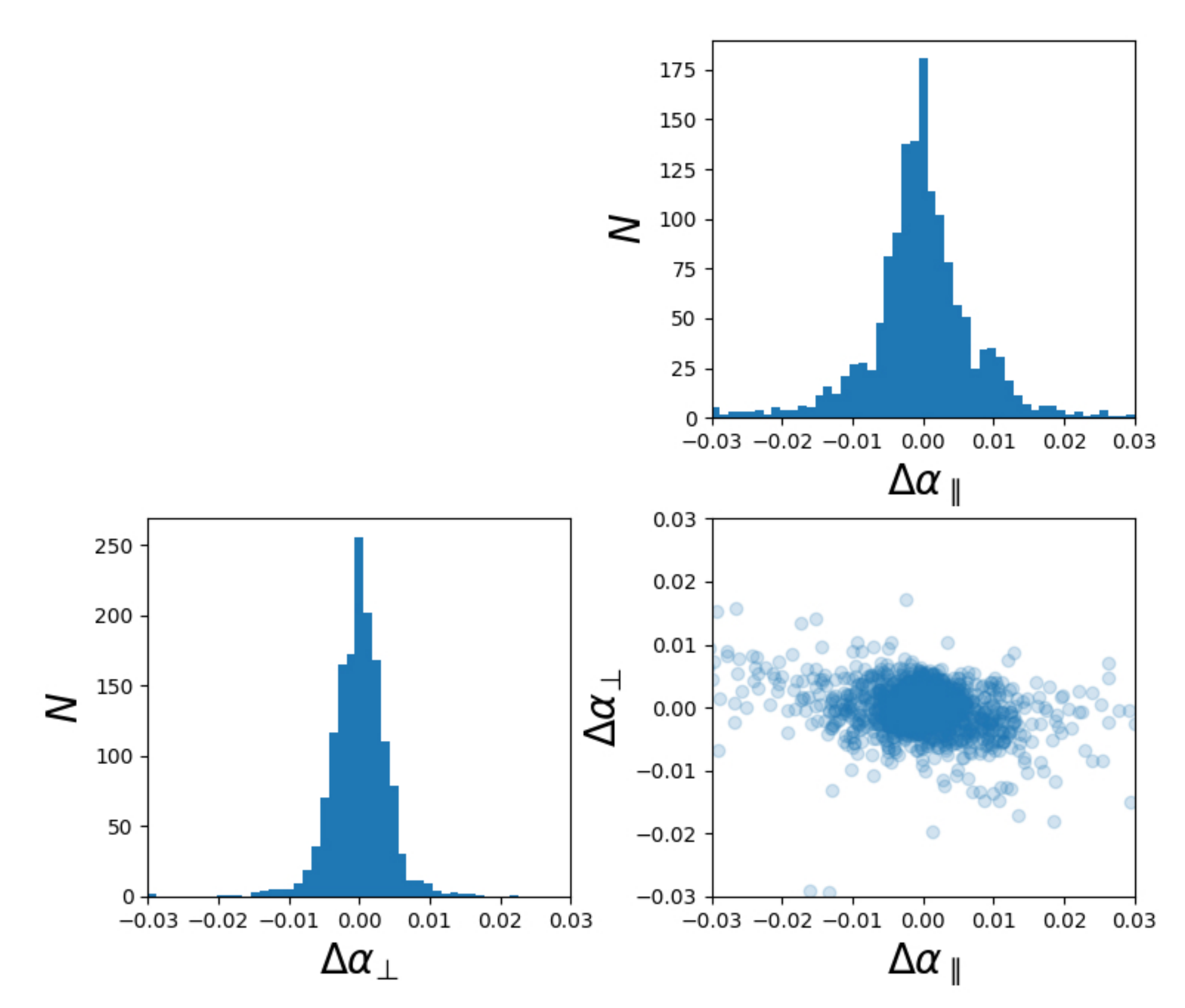}
    \caption{Distributions of measured $\Delta\alpha_{\bot}$ and $\Delta\alpha_{\parallel}$ for the 1600 combinations of true box and assumed cosmologies used during reconstruction and the fitting. These distributions have $\overline{\Delta\alpha_{\bot}}\pm \sigma_{\Delta\alpha_{\bot}} = -0.000047 \pm 0.0051$ and $\overline{\Delta\alpha_{\parallel}}\pm \sigma_{\Delta\alpha_{\parallel}} = -0.00060 \pm 0.010$ both consistent with 0 as expected.}
    \label{fig:delalp_S2}
\end{figure}

The distributions of $\Delta\alpha_{\bot}$ and $\Delta\alpha_{\parallel}$ are shown in Figure~\ref{fig:delalp_S2} with $\overline{\Delta\alpha_{\bot}}\pm \sigma_{\Delta\alpha_{\bot}} = -0.000047 \pm 0.0051$ and $\overline{\Delta\alpha_{\parallel}}\pm \sigma_{\Delta\alpha_{\parallel}} = -0.00060 \pm 0.010$ both consistent with 0 as expected.

The full analysis pipeline steps tested, whilst also allowing for an effective removal of sample variance, is as follows:
\begin{enumerate}
    \item Rescale the coordinate system used when analysing the data to mimic the AP effect.
    \item Apply reconstruction using the assumed fiducial cosmology.
    \item Remove the rescaling to provide a power spectrum with equivalent sample variance.
    \item Generate the power spectrum using the true simulation cosmology.
    \item Apply BAO fitting using a model with the assumed fiducial cosmology.
    \item Translate the measured scale parameters for comparison.
\end{enumerate}

\subsection{Trends as a function of cosmological parameter}

\begin{figure*}
    \centering
	\includegraphics[width=2.1\columnwidth]{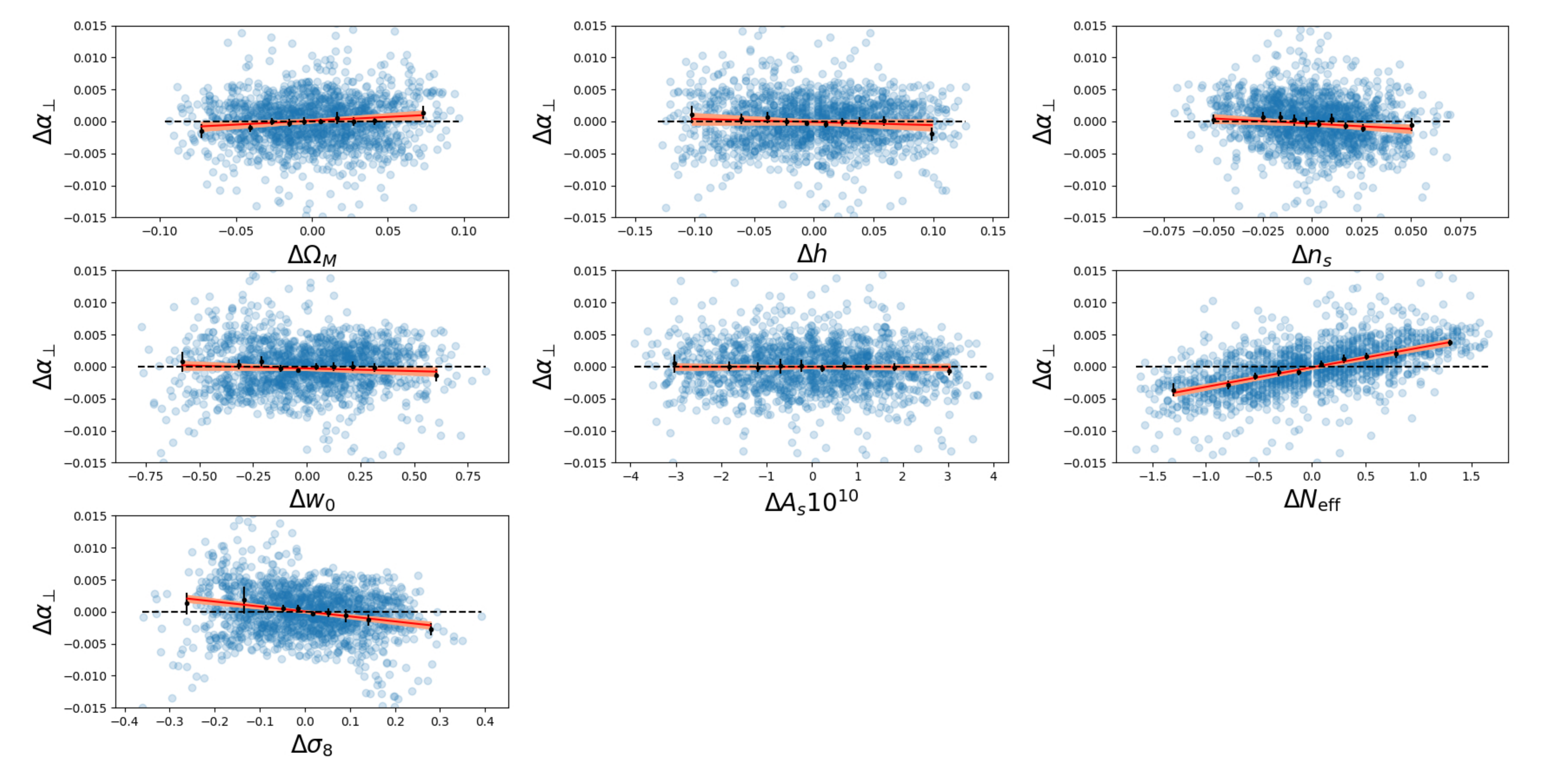}
    \caption{The distribution of $\Delta\alpha_{\bot}$ against the cosmological parameters which have been varied in the {\sc Aemulus} simulations, for tests on the full pipeline. The 1600 scatter points have also been binned in 10 equally sized bins (black data points). The errorbars on these points correspond to the square root of the diagonal elements of the jack-knife re-sample generated covariance matrix. Comparison between the best fit linear trend and a zero-bias flat model show evidence for a trend when varying $N_{\mathrm{eff}}$. In the case of large shifts in cosmology this appears to be a systematic offset of $\sim 0.003$, however for realistic discrepancies between true and fiducial cosmology the bias is negligible.}
    \label{fig:trendS2}
\end{figure*}

\begin{figure*}
    \centering
	\includegraphics[width=2.1\columnwidth]{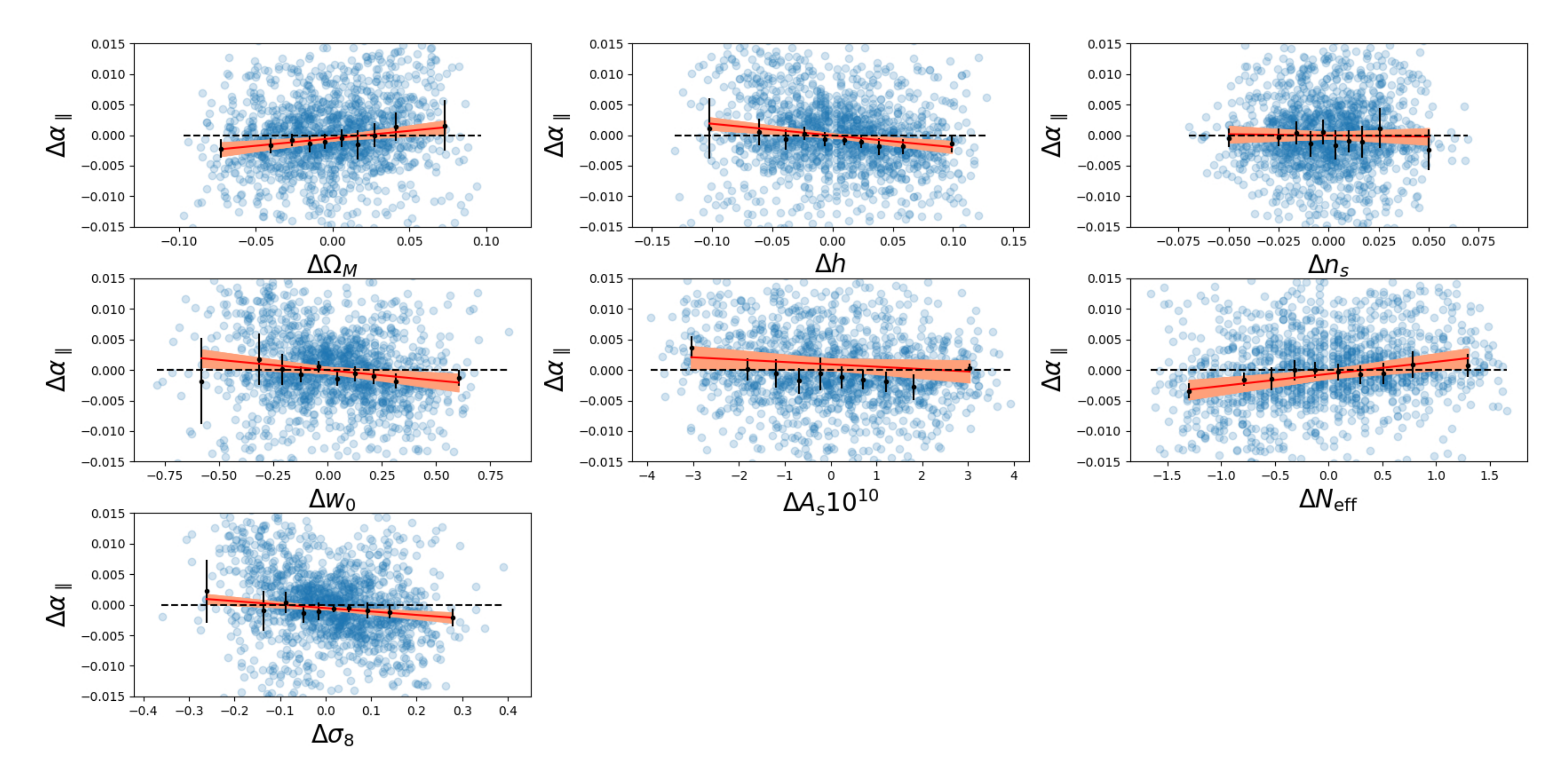}
    \caption{The distribution of $\Delta\alpha_{\parallel}$ against the cosmological parameters which have been varied in the {\sc Aemulus} simulations, for tests on the full pipeline. The 1600 scatter points have also been binned in 10 equally sized bins (black data points). The errorbars on these points correspond to the square root of the diagonal elements of the jack-knife re-sample generated covariance matrix. Comparison between the best fit linear trend and a zero-bias flat model shows mild evidence for deviations when varying $N_{\mathrm{eff}}$. However for realistic discrepancies between true and fiducial cosmology the bias is negligible.}
    \label{fig:trend2S2}
\end{figure*}

Similarly to Section~\ref{sec:trends-recon}, where we considered reconstruction only, we now look for potential biases with respect to changes in our assumed cosmological model for both reconstruction and fitting to the data. The trends for $\Delta\alpha_{\bot}$ and $\Delta\alpha_{\parallel}$ are given in Figures~\ref{fig:trendS2} and~\ref{fig:trend2S2} respectively. In general, for the different fiducial cosmologies assumed now for the entire pipeline, there is a minimal deviation from an unbiased trend. As before, we make a comparison between the best fit linear model and a flat zero bias model to determine the level of evidence. The comparisons in $\Delta AIC$, $\Delta BIC$ and $\sqrt{\Delta\chi^{2}}$ are given in Table~\ref{tab:2}. In almost all cases, even for very large shifts in cosmological parameters, there is no evidence for a non-zero bias in the anisotropic $\alpha$ measured. The exception is $N_{\mathrm{eff}}$ where, for large deviations in the cosmological parameters we can see evidence for a shift at $\Delta\alpha_{\bot} < 0.003$ and $\Delta\alpha_{\parallel} < 0.003$. It should again be stated however that the drivers of this deviation correspond to variations in $N_{\mathrm{eff}}$ of $\pm 1$. For more realistic differences between the true and fiducial cosmology, the shifts are negligible.

\begin{figure*}
    \centering
	\includegraphics[width=1.8\columnwidth]{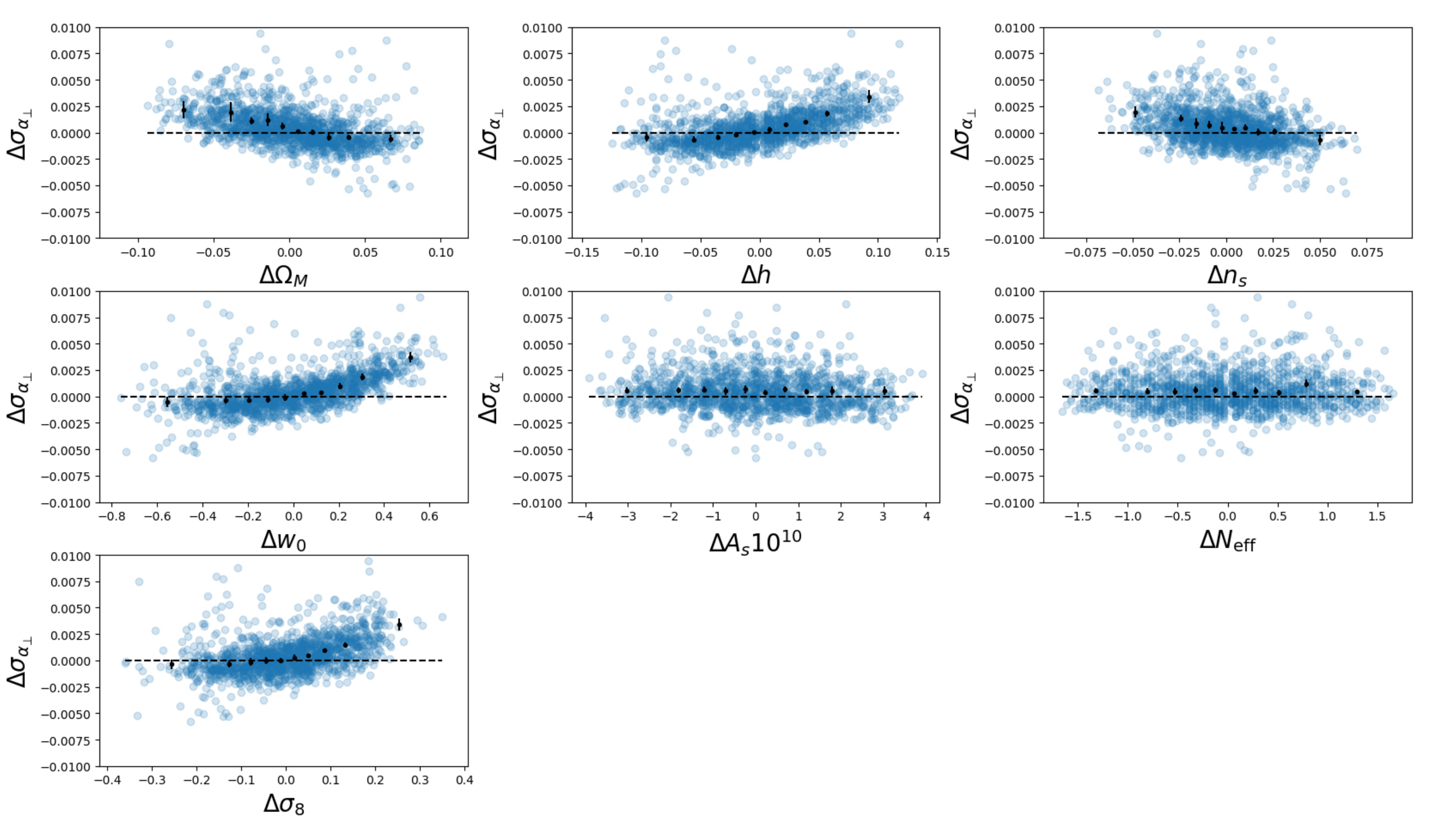}
    \caption{The distribution of $\Delta\sigma_{\alpha_{\bot}}$ against the cosmological parameters which have been varied in the {\sc Aemulus} simulations, for tests on the full pipeline.  The 1600 scatter points have also been binned in 10 equally sized bins (black data points). The errorbars on these points correspond to the square root of the diagonal elements of the jack-knife re-sample generated covariance matrix. There can be seen significant deviations, in the case of large differences between cosmologies, from a zero-bias trend. Suggested evidence for an increase by up to $\Delta\sigma_{\alpha_{\bot}} = +0.002$ when incorrect cosmology leads to inefficent reconstruction.}
    \label{fig:trend3S2}
\end{figure*}

\begin{figure*}
    \centering
	\includegraphics[width=1.8\columnwidth]{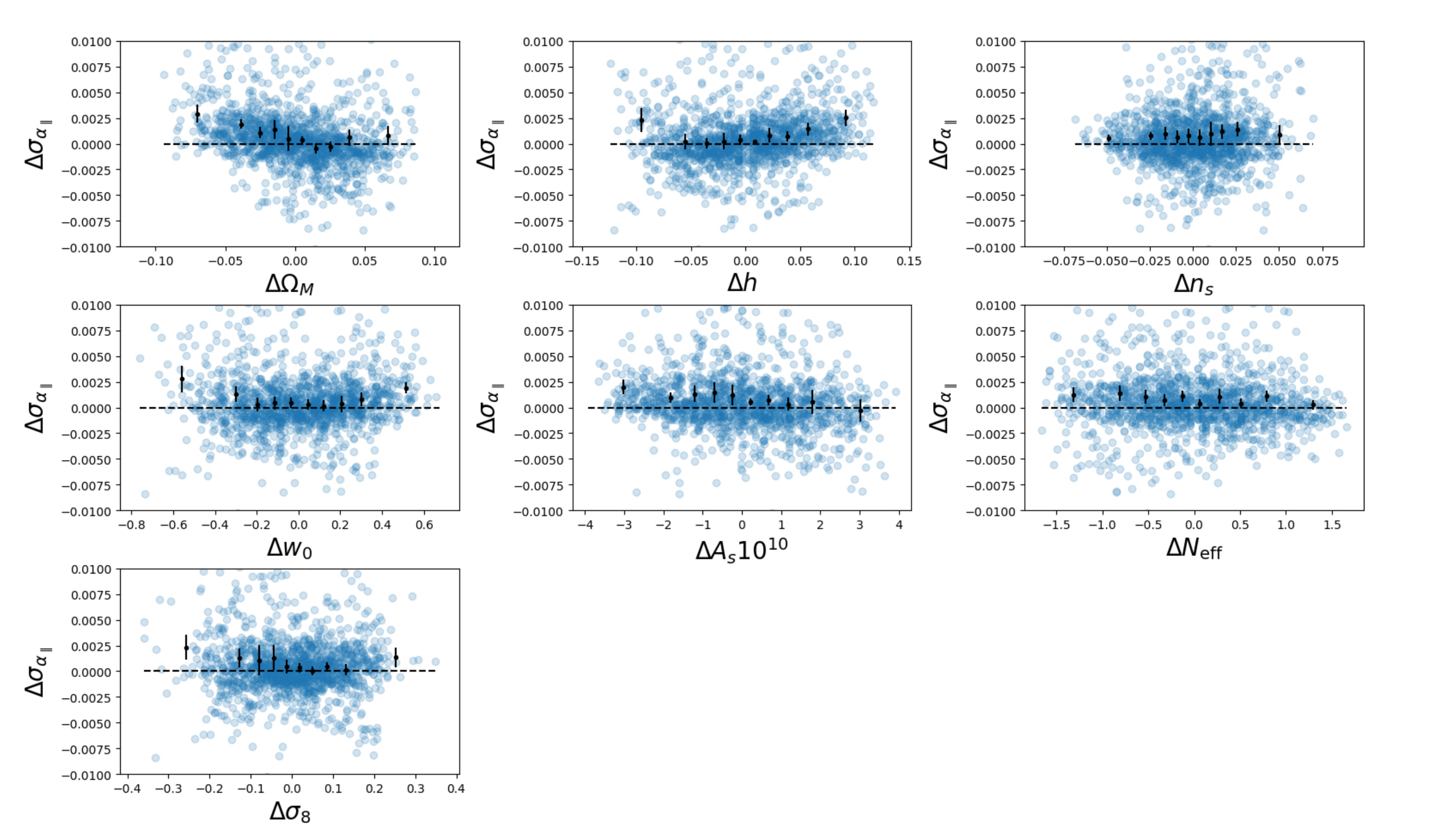}
    \caption{The distribution of $\Delta\sigma_{\alpha_{\parallel}}$ against the cosmological parameters which have been varied in the {\sc Aemulus} simulations, for tests on the full pipeline.  The 1600 scatter points have also been binned in 10 equally sized bins (black data points). The errorbars on these points correspond to the square root of the diagonal elements of the jack-knife re-sample generated covariance matrix. There can be seen significant deviations, in the case of large differences between cosmologies, from a zero-bias trend. Suggested evidence for an increase by up to $\Delta\sigma_{\alpha_{\parallel}} = +0.002$ when incorrect cosmology leads to inefficent reconstruction.}
    \label{fig:trend4S2}
\end{figure*}

We also investigate the trend of $\sigma_{\alpha_{\parallel/\bot}}$ against changes in the cosmological parameters. The trends of $\Delta\sigma_{\alpha_{\parallel/\bot}}$ are shown in Figure~\ref{fig:trend3S2} and~\ref{fig:trend4S2}. As in Section~\ref{sec:trends-recon}, large shifts in fiducial cosmology away from the truth can lead to positive deviations of $\Delta\sigma_{\alpha_{\parallel/\bot}}$. This is expected because optimal reconstruction (which occurs when using the true cosmology) should on average provide the lowest uncertainty on $\alpha$ as it has more effectively removed the non-linear evolution from the galaxy density field.

\begin{table}
 \centering
 \caption{Tabulated results of the model comparison tests between the best fit linear trend and a flat no bias line at $\Delta\alpha = 0$. For each of the 1600 combinations measurements of $\Delta\alpha_{\bot}$ and $\Delta\alpha_{\parallel}$, in the case where we are testing reconstruction and the fitting analysis, we give the AIC, BIC and $\sqrt{\Delta\chi^{2}}$ as complimentary comparison indicators. Minor evidence is shown in some cases of variations in cosmological parameters for $\Delta\alpha_{\bot}$, although these trends are in general driven by extreme shifts in underlying model.}
 \label{tab:2}
 \begin{tabular}{cccccccc}
  \hline
   & & $\alpha_{\bot}$ & & \vline & & $\alpha_{\parallel}$ &\\
  \hline
   & $\Delta \mathrm{AIC}$ & $\Delta \mathrm{BIC}$ & $\sqrt{\Delta\chi^{2}}$ & \vline & $\Delta \mathrm{AIC}$ & $\Delta \mathrm{BIC}$ & $\sqrt{\Delta\chi^{2}}$\\ 
  \hline
  $\Delta \Omega_{m}$ & -1.5 & -2.1 & 1.57 & & 0.4 & -0.2 & 2.10\\
  $\Delta h$ & -3.3 & -3.9 & 0.86 & & -0.5 & -1.1 & 1.88\\
  $\Delta n_{s}$ & 0.7 & 0.1 & 2.17 & & -4.0 & -4.6 & 0.15\\
  $\Delta w_{0}$ & -2.9 & -3.5 & 1.06 & & -0.5 & -1.1 & 1.87\\
  $\Delta A_{s}10^{10}$ & -4.0 & -4.6 & 0.19 & & -1.1 & -1.7 & 1.71\\
  $\Delta N_{\mathrm{eff}}$ & 118.0 & 117.4 & 11.04 & & 8.9 & 8.3 & 3.60\\
  $\Delta \sigma_{8}$ & 5.4 & 4.8 & 3.06 & & 0.1 & -0.5 & 2.03\\
  \hline
 \end{tabular}
\end{table}

\section{Conclusion}  \label{sec:conclusion}

In this study, we have made use of the {\sc Aemulus} suite \citep{2018arXiv180405865D} of $w\mathrm{CDM}$ halo catalogues to test for systematic biases in the measured BAO scale arising from using an incorrect fiducial cosmology during the application of density reconstruction and through the BAO template fitted. We measured the anisotropic scales $\alpha_{\bot}$ and $\alpha_{\parallel}$ in all possible combinations of assumed fiducial and underlying true cosmology for the 40 simulation boxes. This gave a grid of 1600 combinations from which to sample systematic trends against changes in the cosmological model. In order to measure this potential bias to below the level of forecasted precision achievable with future surveys, such as DESI, Euclid and WFIRST \citep{2019arXiv190205569A}, we considered the difference between measurements using the true cosmology for a simulation and with the incorrect assumption, $\Delta\alpha_{\bot}$ and $\Delta\alpha_{\parallel}$. This effectively removes sample variance from the measurement allowing us to reach $\sim 0.3\%$ measurements in $\Delta\alpha_{\bot}$ and $\sim 0.7\%$ in $\Delta\alpha_{\parallel}$ in any one sample. By then binning these in bins with equal numbers of samples the overall trend of the potential systematic can be measured to higher precision.

For the test on how an incorrectly assumed fiducial cosmology affects density field reconstruction only we find, for the overall distribution of all combinations, $\alpha_{\bot}\pm\sigma_{\alpha_{\bot}} = 1.0029\pm0.010$, $\alpha_{\parallel}\pm\sigma_{\alpha_{\parallel}} = 1.0052\pm0.024$ both consistent with $\alpha_{\bot} = \alpha_{\parallel} = 1$ and $\Delta\alpha_{\bot}\pm\sigma_{\Delta\alpha_{\bot}} = 0.00060\pm0.0029$, $\Delta\alpha_{\parallel}\pm\sigma_{\Delta\alpha_{\parallel}} = 0.0013\pm0.0069$ both consistent with $\Delta\alpha_{\bot} = \Delta\alpha_{\parallel} = 0$ as expected. When looking at the trends from the binned $\Delta\alpha_{\bot}$, $\Delta\alpha_{\parallel}$ against cosmology parameters which have been varied between halo catalogues there is no evidence for a systematic bias in almost all cases. Both $\Omega_{M}$ and $h$ have evidence ($\Delta \mathrm{AIC}, \Delta\mathrm{BIC}>5$ and $\sqrt{\Delta\chi^{2}}>3$) for a small $+0.1\%$ systematic bias in $\Delta\alpha_{\bot}$, however it should be noted that these trends appear to be due to deviations from a zero trend in the case of large shifts in cosmological parameters (outside $3\sigma$ from the joint CMB experiments maximum likelihood parameters).

We have also tested how the fiducial cosmology used in reconstruction affects the errors on $\alpha$, as a function of cosmological parameters. We observe an increase in the error on $\alpha_{\bot}$ and $\alpha_{\parallel}$ to a significant level in the case of medium to large shifts in fiducial cosmology away from the truth. The deviations are positive, indicating that in the case where the fiducial cosmology is incorrectly applied one obtains an unbiased value of $\alpha_{\bot}$ and $\alpha_{\parallel}$ but can increase the uncertainty by a factor of $\Delta\sigma_{\alpha_{\bot}} \sim +0.001$ and $\Delta\sigma_{\alpha_{\parallel}} \sim +0.002$ for reasonable shifts (within $3\sigma$ of the joint CMB experiments Likelihood).

For incorrect assumptions of fiducial cosmology in the density field reconstruction procedure only allowed by current experiments, we have shown that there is negligible induced bias in the measurements of $\alpha_{\bot}$ and $\alpha_{\parallel}$, at the level soon to be probed by future surveys. This is consistent with results recovered from theoretical modelling \citep{2019JCAP...02..027S} in which they find $10^{-4}$ shifts for small differences in the cosmological model ($3\%$ errors in the distance). However, since our analysis is using the standard BAO analysis pipeline, our results represent a necessary step to connect the theory explored in \cite{2019JCAP...02..027S} with data analysis.

The assumption of incorrect fiducial cosmology influences more in the analysis pipeline than just reconstruction. The reference frame that the power spectrum is measured in has been distorted when transforming from observed to cartesian coordinates and also the underlying linear power spectrum used in the modelling template is generated using this cosmology. We have tested the way that the fiducial cosmology can affect the entire BAO pipeline by repeating the above analysis without isolating the assumptions adopted within the reconstruction algorithm only.

For the full pipeline test we find $\Delta\alpha_{\bot}\pm\sigma_{\Delta\alpha_{\bot}} = -0.000047\pm0.0051$, $\Delta\alpha_{\parallel}\pm\sigma_{\Delta\alpha_{\parallel}} = -0.00060\pm0.010$ both consistent with $\Delta\alpha_{\bot} = \Delta\alpha_{\parallel} = 0$. When looking at trends with cosmological parameter there is  negligible deviation with the exception of changes in $N_{\mathrm{eff}}$. However for any reasonably expected differences between the true and assumed cosmology the bias is $< 0.1\%$. Similarly to the reconstruction only case, significant increases in $\sigma_{\alpha_{\parallel}}$ and $\sigma_{\alpha_{\bot}}$ are seen in the case of large differences between the truth and assumed cosmologies. This correlates well with what is seen in the reconstruction only case with potential increases of $\sigma_{\alpha_{\bot}} \sim 0.1\%$ and $\sigma_{\alpha_{\parallel}} \sim 0.2\%$ for differences within $3\sigma$ of CMB constraints.

The results of this paper are consistent with what has been seen previously both from theoretical studies and smaller survey specific investigations. The implications for future surveys are that we see no evidence for any additional systematic error budget in measured $\alpha_{\parallel}$ and $\alpha_{\bot}$ to $< 0.1\%$. 

\section*{Acknowledgements}
PC acknowledges support from the European Research Council through the Dark survey grant 614030. FB is a Royal Society University Research Fellow. Some of the simulations for this project were performed on the Sherlock cluster. We would like to thank Stanford University and the Stanford Research Computing Center for providing computational resources and support that contributed to these research results. JD and RHW thank Matt Becker and their collaborators in the {\sc Aemulus} collaboration for their contributions to the simulation suite used in this work. This research used resources of the National Energy Research Scientific Computing Center (NERSC), a U.S. Department of Energy Office of Science User Facility operated under Contract No. DE-AC02-05CH11231. JD and RHW received partial support from the U.S. Department of Energy under contract number DE-AC02-76SF00515.



\bibliographystyle{mnras}




\bsp	
\label{lastpage}
\end{document}